\documentclass[12pt]{article}
\usepackage[dvips]{graphicx}
\usepackage{epsfig}
\psfigdriver{dvips}

\renewcommand{\thefootnote}{\fnsymbol{footnote}}

\newcommand{\prepr}[1] {\begin{flushright}  {\bf #1} \end{flushright} \vskip 1.cm}
\newcommand{\titul}[1] {\boldmath \begin{center}{\Large {\bf #1 } } \end{center}
\vskip 0.8cm}
\newcommand{\autor}[1] {\begin{center}  {\bf \lineskip .3cm #1  }
                        \end{center} }
\newcommand{\lugar}[1] {\begin{center}  {\normalsize \bf \it #1   } \end{center}}

\newcounter{muni}

\def\bmaT{\left(\begin{array}{ccc}}
\def\emaT{\end{array}\right)}
\def\bma{\left( \begin{array} }
\def\ema{\end{array} \right)}
\def\gsim{~{\rlap{\lower 3.5pt\hbox{$\mathchar\sim$}}\raise 1pt\hbox{$>$}}\,}
\def\lsim{~{\rlap{\lower 3.5pt\hbox{$\mathchar\sim$}}\raise 1pt\hbox{$<$}}\,}

\makeatletter
\def\fmslash{\@ifnextchar[{\fmsl@sh}{\fmsl@sh[0mu]}}
\def\fmsl@sh[#1]#2{%
  \mathchoice
    {\@fmsl@sh\displaystyle{#1}{#2}}%
    {\@fmsl@sh\textstyle{#1}{#2}}%
    {\@fmsl@sh\scriptstyle{#1}{#2}}%
    {\@fmsl@sh\scriptscriptstyle{#1}{#2}}}
\def\@fmsl@sh#1#2#3{\m@th\ooalign{$\hfil#1\mkern#2/\hfil$\crcr$#1#3$}}
\makeatother
\begin{document}
\hbadness=10000
\pagenumbering{arabic}
\begin{titlepage}
\prepr{hep-ph/0506176\\
\hspace{30mm} KEK--TH--1023 \\
\hspace{30mm} June 2005}

\begin{center}
\titul{\bf Single and Pair Production of Doubly Charged\\
Higgs Bosons at Hadron Colliders}

\autor{A.G. Akeroyd\footnote{akeroyd@post.kek.jp} and 
Mayumi Aoki\footnote{mayumi.aoki@kek.jp}}
\lugar{KEK Theory Group,\\
1-1 Oho \\
Tsukuba, Japan 305-0801}

\end{center}

\vskip2.0cm

\begin{abstract}
\noindent
Current searches for doubly charged Higgs bosons 
($H^{\pm\pm}$) at the Fermilab Tevatron
are sensitive to single production of $H^{\pm\pm}$,
although the pair production mechanism
$q\overline q\to H^{++}H^{--}$ is assumed to be dominant.
In the context of a Higgs Triplet Model we study 
the mechanism $q'\overline q\to H^{\pm\pm}H^{\mp}$
at the Tevatron and CERN Large Hadron Collider,
and show that its inclusion can significantly 
improve the search potential for $H^{\pm\pm}$.
Moreover, assuming that the neutrino mass is generated solely by the
triplet field Yukawa coupling to leptons, we compare the branching 
ratios of $H^{\pm\pm}\to l^\pm l^\pm$ and $H^{\pm\pm}\to H^\pm W^*$ for 
the cases of a normal hierarchical, inverted hierarchical and 
degenerate neutrino mass spectrum.

\end{abstract}

\vskip1.0cm
{\bf  PACS index :12.60.Fr,14.80.Cp,14.60.Pq}
\end{titlepage}
\thispagestyle{empty}
\newpage

\pagestyle{plain}
\renewcommand{\thefootnote}{\arabic{footnote} }
\setcounter{footnote}{0}

\section{Introduction}
The quest for Higgs bosons is of utmost importance at high energy
colliders \cite{Gunion:1989we},\cite{Carena:2002es},
\cite{Djouadi:2005gi}.
In the Standard Model (SM), one isospin $I=1/2$, 
hypercharge $Y=1$ complex scalar doublet breaks the 
Electroweak Symmetry (EW) and provides mass for the fermions, $W^\pm$ 
and $Z$. One neutral scalar, $\phi^0$, remains
as a physical degree of freedom -- ``the SM Higgs boson''.
Such a framework predicts $\rho\,(=M^2_W/M^2_Z\cos^2\theta_W)=1$ 
at tree-level, 
a result which is in impressive agreement with
the experimental measurement of $\rho\approx 1$
\cite{Eidelman:2004wy}.  
More generally, any Higgs sector composed solely of $I=1/2, Y=1$ doublets 
assures $\rho=1$ at tree-level, with calculable 1-loop corrections
\cite{Hollik:1986gg}.

Predicting $\rho=1$ at tree-level is certainly an
attractive feature of $I=1/2,Y=1$ doublet representations,
although models with isospin triplets ($I=1$) 
can also be considered \cite{Gunion:1989we}. Such models have 
various virtues and
deficiencies. If the neutral member of the triplet acquires a 
vacuum expectation value (VEV) then
$\rho=1$ at tree-level is no longer guaranteed, and the 
triplet VEV must be very small in order to comply with the
measured value $\rho\approx 1$. However, unlike doublets, 
$Y=2$ triplets can give rise to neutrino masses and mixings 
whose magnitude is proportional to the triplet vacuum expectation value 
multiplied by an arbitrary Yukawa coupling ($h_{ij}$)
{\sl without} invoking a right handed neutrino \cite{Schechter:1980gr},
\cite{Gelmini:1980re}.

A clear phenomenological signature of $Y=2$ triplets 
would be the observation of a doubly charged Higgs boson
$H^{\pm\pm}$. Such $H^{\pm\pm}$ have been searched for at 
the $e^+e^-$ collider LEP, resulting in mass
limits of the order $m_{H^{\pm\pm}} > 100$ GeV
\cite{Abdallah:2002qj},\cite{Abbiendi:2001cr},
\cite{Achard:2003mv},\cite{Abbiendi:2003pr}. 
Their existence can also affect a wide variety of
processes, such as Bhabha scattering, the anomalous 
magnetic moment of the muon $(g-2)_\mu$, and 
lepton flavour violating $\mu^\pm$ and $\tau^\pm$ decays 
\cite{Swartz:1989qz},\cite{Cuypers:1996ia},\cite{Chun:2003ej},
\cite{Kakizaki:2003jk},\cite{Atag:2003wk},\cite{Boyarkin:2004zd}.
The Fermilab Tevatron recently performed the first search for 
$H^{\pm\pm}$ at hadron colliders. 
The production process $p\overline p\to \gamma,Z\to H^{++}H^{--}$
was assumed, with subsequent decay $H^{\pm\pm}\to l^\pm l^\pm$.
D0 \cite{Abazov:2004au}
searched for $H^{\pm\pm}\to \mu^\pm\mu^\pm$ while CDF
\cite{Acosta:2004uj}
searched for 3 final states 
$H^{\pm\pm}\to \mu^\pm\mu^\pm,\mu^\pm e^\pm$ and
$e^\pm e^\pm$. Mass limits of the order
$m_{H^{\pm\pm}}> 130$ GeV were obtained
with an integrated luminosity of 240 pb$^{-1}$,
assuming BR($H^{\pm\pm}\to l^\pm_i l^\pm_j)=100\%$ \cite{Acosta:2004uj}
in a given channel. These are 
the strongest direct mass limits on any type of Higgs boson,
which shows the strong search capability of hadron colliders in the
channel $H^{\pm\pm}\to l^\pm l^\pm$. 

Given this strong search potential, in this paper we
consider the phenomenological effect of relaxing these 
simplifying assumptions for the dominant
production mechanism and decay modes of $H^{\pm\pm}$.
Although work along these lines has appeared previously
\cite{Chun:2003ej},
\cite{Gunion:1989in},\cite{Gunion:1996pq},\cite{Chakrabarti:1998qy},
\cite{Gunion:1998ii},\cite{Dion:1998pw}
we develop and expand the preceding analyses. 
For example, if $h_{ij}$ are solely responsible
for the currently favoured form of the neutrino mass matrix then
BR($H^{\pm\pm}\to l^\pm_i l^\pm_j)< 100\%$ 
in a given channel \cite{Chun:2003ej}.
In this paper we study in detail the alternative 
production mechanism $q'\overline q\to W^*\to H^{\pm\pm} H^\mp$
\cite{Dion:1998pw}, 
which can be as large as $q\overline q\to \gamma,Z\to H^{++}H^{--}$. 
Since the current search strategy at the Tevatron 
is in fact sensitive to {\sl single} production of
$H^{\pm\pm}$, we introduce 
the inclusive single production cross-section
$(\sigma_{H^{\pm\pm}}$) as the sum of the single and pair production
cross-sections. We point out that the contribution of
$q'\overline q\to W^*\to H^{\pm\pm} H^\mp$ to $\sigma_{H^{\pm\pm}}$
strengthens the Tevatron mass limit on $H^{\pm\pm}$, which in general
has a dependence on $m_{H^\pm}$.
Moreover, we quantify the impact of the potentially important decay mode 
$H^{\pm\pm}\to H^\pm W^*$ \cite{Chakrabarti:1998qy}
in the light of recent neutrino data. Although such a decay 
can weaken the $H^{\pm\pm}$ search capability in the leptonic channel,
observation of $H^{\pm\pm}\to H^\pm W^*$ together with
one or more leptonic channels might permit an
order of magnitude estimate of $h_{ij}$ \cite{Gunion:1996pq},
\cite{Gunion:1998ii}.

Our work is organized as follows.
In Section 2 we introduce the Higgs Triplet Model. In Section 3
we study the production mechanism 
$q'\overline q\to H^{\pm\pm} H^\mp$ and its phenomenological
effect on the $H^{\pm\pm}$ search at the Tevatron and LHC.
In Section 4 we quantify the impact of the decay 
$H^{\pm\pm}\to H^\pm W^*$,
while Section 5 considers the search potential of the
Tevatron in the generalized scenario. 
Finally, in Section 6 we present our conclusions.

\section{The Higgs Triplet Model}
Higgs $I=1$ triplet representations arise in several
well motivated models of physics beyond the SM 
\cite{Gunion:1989we},\cite{Gunion:1989in}.
For example, Left-Right (L-R) symmetric models built on the gauge group
$SU(2)_R\times SU(2)_L \times U(1)$ contain both left- and 
right-handed $I=1,Y=2$ triplet representations. Such models 
also require extra gauge bosons and can provide naturally
light neutrino masses via the seesaw mechanism. 
Little Higgs models \cite{Arkani-Hamed:2001nc} also
require $I=1,Y=2$ triplet representations,
as well as new gauge bosons and fermions.
However, Higgs triplets can be considered 
as a minimal addition to the SM 
\cite{Gunion:1989ci}-
for a review see \cite{Kundu:1995qb}.
We will focus on a particularly simple model \cite{Schechter:1980gr},
\cite{Gelmini:1980re} which merely adds
a $I=1,Y=2$ complex (left-handed) Higgs triplet to the SM Lagrangian, 
hereafter referred
to as the ``Higgs Triplet Model'' or ``HTM''.
Such a model can provide a Majorana mass for the observed neutrinos 
without the need for a right handed neutrino via the 
gauge invariant Yukawa interaction:
\footnote{Note that the analogous term for a $Y$=0 triplet 
is forbidden by gauge invariance}
\begin{equation}
{\cal L}=h_{ij}\psi_{iL}^TCi\tau_2\Delta\psi_{jL}+h.c
\end{equation}
Here $h_{ij} (i,j=1,2,3)$ is an arbitrary coupling,
$C$ is the Dirac charge conjugation operator, 
$\psi_{iL}=(\nu_i, l_i)_L^T$ is a left-handed lepton doublet,
and $\Delta$ is a $2\times 2$ representation of the $Y=2$
complex (left-handed) triplet fields:
\begin{equation}
\Delta
=\bma{cc}
\Delta^+/\sqrt{2}  & \Delta^{++} \\
\Delta^0       & -\Delta^+/\sqrt{2}
\ema
\end{equation}
The Higgs potential \cite{Chun:2003ej}
is as follows, with $\Phi=(\phi^+,\phi^0)^T$:
\begin{eqnarray}
V&=&m^2(\Phi^\dagger\Phi)+\lambda_1(\Phi^\dagger\Phi)^2+M^2
{\rm Tr}(\Delta^\dagger\Delta) +
\lambda_2[{\rm Tr}(\Delta^\dagger\Delta)]^2+ \lambda_3{\rm Det}
(\Delta^\dagger\Delta)  \nonumber \\
&&+\lambda_4(\Phi^\dagger\Phi){\rm Tr}(\Delta^\dagger\Delta)
+\lambda_5(\Phi^\dagger\tau_i\Phi){\rm Tr}(\Delta^\dagger\tau_i
\Delta)+\left(
{1\over \sqrt 2}\mu(\Phi^Ti\tau_2\Delta^\dagger\Phi) + h.c \right)
\end{eqnarray}
The term $\mu \Phi\Delta\Phi$, where $\mu$ is
a dimensionful trilinear coupling, gives rise to
a VEV $v_\Delta$ for the neutral member of the triplet
$\Delta^0$:
\begin{equation}
v_\Delta \simeq \mu v^2/2M^2
\end{equation} 
Here $M$ is the common triplet mass ($M^2\Delta^\dagger\Delta$).
Since we are interested in the
case of light triplets we take $M\approx v$, and so 
$v_\Delta\approx \mu$.
A non-zero $v_\Delta$ gives rise to the following mass matrix for 
neutrinos:
\begin{equation}
m_{ij}=2h_{ij}\langle\Delta^0\rangle = \sqrt{2}h_{ij}v_{\Delta}
\label{nu_mass}
\end{equation}
Note that the HTM is free from a massless Goldstone boson
(Majoron) 
arising from the violation of the lepton number ($L$) global symmetry,
because the Higgs potential contains the term 
$\mu\Phi\Delta\Phi$ term which explicitly violates lepton number
when $\Delta$ is assigned $L=-2$. 
Cosmological data provides a constraint on the neutrino masses
$m_i$, $\Sigma m_i \lsim 0.75$ eV \cite{Barger}. 
Lepton flavour 
violating (LFV) processes involving 
$\mu$ and $\tau$
provide the strongest upper limits on $h_{ij}$ and hence 
$v_\Delta$ cannot be
arbitrarily small if the HTM is to accommodate the currently 
favoured form
of the neutrino mass matrix. A rough lower bound $v_\Delta \gsim 10$ eV 
can be derived. An upper limit on $v_\Delta$ can be obtained from
considering its effect on $\rho$. In the HTM $\rho$ is given
by (where $x=v_\Delta/v$):
\begin{equation}
\rho\equiv 1+\delta\rho={1+2x^2\over 1+4x^2}
\label{deltarho}
\end{equation}
From the measurement of $\rho\approx 1$
a purely tree-level analysis gives
the bound $v_\Delta/v\lsim 0.03$. 
We will comment on the 1-loop expression for $\delta\rho$
below \cite{Czakon:1999ue},\cite{Chen:2003fm},\cite{Blank:1997qa}.
In this paper we will assume
\begin{equation}
10 ~{\rm eV} \lsim v_\Delta \lsim 10000 ~{\rm eV}
\end{equation}
Hence the tree-level value of $\rho$ is essentially
equal to 1, thus easily satisfying the experimental constraint on
$\delta\rho$.
Such small values of $v_\Delta$ can be explained by a 2 loop
mechanism \cite{Chun:2003ej} or in the context of 
extra dimensions \cite{Ma:2000wp},\cite{Chen:2005mz}. 
Moreover, such values of $v_\Delta$
would permit some $h_{ij}$ to be sufficiently large to enhance 
various LFV $\mu$ and $\tau$ decays to the sensitivity of
current and forthcoming experiments \cite{Chun:2003ej},
\cite{Kakizaki:2003jk},\cite{Boyarkin:2004zd}, and are
also consistent with the requirement that any primordially 
generated baryon asymmetry is not erased by the lepton number 
violating triplet interactions \cite{Hasegawa:2004bx}.

The HTM has 7 Higgs bosons $(H^{++},H^{--},H^+,H^-,H^0,A^0,h^0)$.
While $H^{\pm\pm}$ is purely triplet ($=\Delta^{\pm\pm}$), 
the remaining eigenstates would in general be mixtures of the  
doublet and triplet fields. Such mixing is proportional to the 
triplet VEV, and hence small {\it even if} $v_\Delta$
assumes its largest value of a few GeV.
Therefore the first six eigenstates are essentially composed 
of triplet fields, while the $I=1/2$ doublet gives rise to
a SM like $h^0$ and the Goldstone bosons $G^\pm,G^0$. 
The most striking signature of the HTM
would be the observation of $H^{\pm\pm}$.
\footnote{The dominantly triplet eigenstates $H^{\pm}$, $H^0$ and $A^0$
can have a different phenomenology to the analogous Higgs bosons in 
doublet ($I=1/2, Y=1)$ representations.}
In the HTM there exists the following relationships among the 
masses of the physical Higgs bosons:

\begin{eqnarray}
m^2_{H^{\pm\pm}}\simeq
M^2+2\frac{(\lambda_4-\lambda_5)}{g^2}M^2_W \\ \nonumber
m^2_{H^{\pm}}\simeq
m^2_{H^{\pm\pm}}+2\frac{\lambda_5}{g^2} M^2_W \\ \nonumber
m^2_{H^0,A^0}\simeq
m^2_{H^{\pm}}+2\frac{\lambda_5}{g^2}M^2_W  
\end{eqnarray}

Here $M$ is the triplet mass term, while
$\lambda_4,\lambda_5$ are dimensionless quartic couplings.
For $\lambda_5 > 0$ ($\lambda_5 < 0$)
one has the following hierarchy $m_{H^{\pm\pm}}< m_{H^\pm}
<m_{H^0,A^0}$ ($m_{H^{\pm\pm}}> m_{H^\pm} > m_{H^0,A^0}$).
Clearly $M$ sets the scale for the mass of the triplet fields, 
while the mass splitting among the eigenstates is determined by 
the quartic couplings and can be ${\cal O} (M_W)$.
We will focus on Higgs boson masses of 
interest for the Tevatron and LHC, and hence we assume $M \gsim 1$ TeV.

At the 1-loop level the Higgs sector contribution to 
$\delta\rho$ is a function of $v_{\Delta}$ and 
the Higgs boson masses.
Although a quantitative analysis in the context of the HTM is still 
lacking, explicit formulae for the contributions of 
$Y=2$ triplets to the self-energies of the $W$ and $Z$ 
in the context of 
L-R symmetric models and Little Higgs Models
can be found in \cite{Czakon:1999ue},
\cite{Chen:2003fm}.
In particular,
such contributions are sensitive to the mass splittings
of the Higgs bosons. In the HTM
the triplet Higgs boson mass splitting is determined by 
the quartic coupling $\lambda_5$, with $\lambda_5=0$
giving rise to degenerate triplet scalars of mass $M$.
We will present results for both the degenerate case
and for mild splittings of up to 20 GeV in our discussion
of $H^{\pm\pm}$ phenomenology at the Tevatron.

We now briefly discuss present mass bounds on the Higgs bosons of the 
HTM, which differ in some cases from the commonly quoted mass bounds
in the 2HDM. 
If $H^0$ and $A^0$ were the lightest, they could have been produced
at LEP via the mechanism $e^+e^-\to A^0H^0$ (note that $e^+e^-\to ZH^0$ 
is proportional to $v_{\Delta}$ and hence negligible).
However, since $A^0$ and $H^0$ would both decay 
invisibly to $\nu\overline \nu$, Ref. \cite{Datta:1999nc} 
suggested using LEP data
on $\gamma\nu\overline \nu$ (where $\gamma$ arises from
bremsstrahlung from $e^+$ or $e^-$)
and derived the mass limit $m_{{H^0},{A^0}}\gsim 55$ GeV.
Concerning $H^\pm$, LEP searched for
$H^\pm\to cs$ or $\tau\nu_{\tau}$ which are expected 
to be the dominant decays in doublet models, and obtained 
mass limits around $m_{H^\pm}\gsim 80$ GeV.
For the triplet $H^\pm$ 
the decays $H^\pm\to e^\pm\nu,\mu^\pm\nu$ may have large BRs.
However, in this scenario one could presumably use data
from slepton searches $e^+e^-\to \tilde l^+\tilde l^-\to l^+l^-
\chi^0\overline \chi^0$ to derive similar mass limits ($\gsim 80$ GeV)
\cite{Abdallah:2003xe}.
A recent quantitative analysis of the above decays
in the context of a Little Higgs Model can be found in
\cite{Han:2005nk}.
 
Concerning $H^{\pm\pm}$, LEP searched for both left-handed 
$H^{\pm\pm}_L$ and right-handed $H^{\pm\pm}_R$ (which we will
not consider in this paper) via several mechanisms:

\begin{itemize}
\item[{(i)}]
Pair production via $e^+e^-\to \gamma^*,Z^*\to H^{++}H^{--}$
followed by decay to $l^+l^+l^-l^-$ ($l^\pm=e^\pm,\mu^\pm,\tau^\pm$); 
the cross-section is determined by gauge couplings and leads to mass
limits of $m_{H^{\pm\pm}}> 100$ GeV \cite{Abdallah:2002qj},
\cite{Abbiendi:2001cr},\cite{Achard:2003mv}.

\item[{(ii)}] Single production of 
$H^{\pm\pm}$ via $e^+e^-\to H^{\pm\pm}e^\mp
e^\mp$; the rate is determined by the coupling $h_{11}$ and leads
to excluded regions in the plane ($h_{11}, m_{H^{\pm\pm}}$), with
sensitivity up to $m_{H^{\pm\pm}}\lsim 180$ GeV. Limits of
$10^{-2}\to 10^{-1}$ were set on $h_{11}$
\cite{Abbiendi:2003pr}.
\item[{(iii)}]
The effect of $H^{\pm\pm}$ on Bhabha scattering $e^+e^-\to e^+e^-$;
as in (ii) above this leads to excluded regions in the plane 
($h_{11}, m_{H^{\pm\pm}}$) \cite{Achard:2003mv},\cite{Abbiendi:2003pr}
with sensitivity up to $m_{H^{\pm\pm}}\lsim 2$ TeV.
Limits of $10^{-2}\to 10^{-1}$ were set on $h_{11}$.

\end{itemize}
The direct searches for $H^{\pm\pm}$ will 
continue at the hadron colliders, Tevatron and LHC.

\section{Production of $H^{\pm\pm}$ at the Tevatron}

A distinct signature of $H^{\pm\pm}$ would be a pair of same sign 
charged leptons ($e^\pm$ or $\mu^\pm$) with high invariant mass. 
At hadron colliders such a signal has a relatively high 
detection efficiency and 
enjoys essentially negligible background from Standard Model
processes. Earlier theoretical studies of the
search potential for $H^{\pm\pm}$ at such colliders can be 
found in \cite{Gunion:1989in},\cite{Gunion:1996pq}, with a
recent analysis at the LHC in \cite{Azuelos:2005uc}.
The decays of $H^{\pm\pm}$ to states 
involving $\tau^\pm$ are more problematic at hadron colliders, 
although simulations in these channels
\cite{Gunion:1996pq}, \cite{Azuelos:2005uc}
promise sensitivity to values of
$m_{H^{\pm\pm}}$ beyond the LEP limits.
The decays $H^{\pm\pm}\to W^\pm W^\pm$ are proportional to 
$v_{\Delta}$, and can be neglected in the case of 
very small $v_{\Delta}$ of interest to us.

In 2003 the Tevatron performed the first search for $H^{\pm\pm}$
at a hadron collider. D0 \cite{Abazov:2004au} 
have searched for $H^{\pm\pm}\to \mu^+\mu^-$ 
while CDF \cite{Acosta:2004uj}
searched for 3 final states: $H^{\pm\pm}\to e^\pm e^\pm,
e^\pm \mu^\pm, \mu^\pm\mu^\pm$. The assumed production mechanism for
$H^{\pm\pm}$ is $q\overline q\to \gamma^*,Z^*\to H^{++}H^{--}$. 
\footnote{The model-dependent 
contribution from any $Z'$ (which can enhance the
cross-section \cite{Dion:1998pw},\cite{Azuelos:2005uc})
is currently not considered.}
This cross-section depends on only one unknown 
parameter, $m_{H^{\pm\pm}}$, and importantly is not suppressed
by any small factor such as a Yukawa coupling $h_{ij}$ or a triplet VEV.
The search assumes that $H^{\pm\pm}$ is sufficiently long-lived
to decay in the detector, which corresponds to $h_{ll}> 10^{-5}$.
A search for a long lived $H^{\pm\pm}$ decaying
outside the detector 
has been performed in \cite{Acosta:2005np}.
The cross-section also depends on the hypercharge of the
Higgs representation, which is $Y=2$ in the HTM. This
value of $Y$ is also assumed in the experimental searches.
The explicit partonic cross-section at leading order (LO) is as follows
(where $q=u,d$):
\begin{equation}
\sigma_{LO}(q\overline q\to H^{++}H^{--})=
{\pi\alpha^2\over 9Q^2}\beta_1^3\left[e_q^2e^2_H+
{e_q e_H {\rm v}_q{\rm v}_H(1-M_Z^2/Q^2)+
({\rm v}^2_q+{\rm a}_q^2){\rm v}^2_H\over
(1-M_Z^2/Q^2)^2+M_Z^2 \Gamma^2_Z/Q^4} \right]
\label{pairH++}
\end{equation}
Here ${\rm v}_q=(I_{3q}-2e_qs^2_W)/(s_Wc_W)$,
${\rm a}_q=I_{3q}/(s_Wc_W)$, and ${\rm v}_H=(I_{3H}-e_H s^2_W)
/(s_Wc_W)$. The third isospin component is denoted by
$I_{3q}$ ($I_{3H}$) and 
$e_q(e_H)$ is the electric charge of the quark $q$
$(H^{\pm\pm})$. $s_W$ and $c_W$ are $\sin\theta_W$ and $\cos\theta_W$
respectively. $Q^2$ is the partonic centre-of-mass (COM) energy. 
$\alpha$ is the QED coupling evaluated at the scale $Q$, $M_Z$ is the
$Z$ boson mass, $\Gamma_Z$ is the $Z$ boson width, and
$\beta_1=\sqrt{1-4m^2_{H^{\pm\pm}}/Q^2}$.
Order $\alpha_s$ QCD corrections modify the LO cross-section by a factor
$K\approx 1.3$ at the Tevatron 
for $m_{H^{\pm\pm}}<200$ GeV, and $K\approx 1.25$ at the LHC
for $m_{H^{\pm\pm}}<1000$ GeV \cite{Muhlleitner:2003me}.
We neglect the gluon-gluon fusion ($\alpha^2_s$) contribution to
$H^{++}H^{--}$ production, which has no compensatory enhancement 
factor analogous to the $\tan^4\beta$ term for doublet $H^\pm$ production
via $gg\to H^+H^-$ \cite{Willenbrock:1986ry}.

Assuming that $H^{\pm\pm}$ production proceeds via this pair
production process, the absence of signal enables a 
limit to be set on the product:
\begin{equation}
\sigma(p\overline p\to H^{++}H^{--})\times BR(H^{\pm\pm}\to l^\pm_il^\pm_j)
\label{crossBR}
\end{equation}

Clearly the strongest constraints on $m_{H^{\pm\pm}}$ are obtained
assuming BR$(H^{\pm\pm}\to l^\pm_il^\pm_j)=100\%$.
Currently these mass limits stand at:
133,115,136 GeV for the $e^\pm e^\pm,e^\pm\mu^\pm,
\mu^\pm\mu^\pm$ channels respectively \cite{Acosta:2004uj}. 
In the HTM one expects $BR(H^{\pm\pm}\to l^\pm_il^\pm_j)
\ne 100\%$ if Eqn.(\ref{nu_mass})
is required to explain the currently favoured form 
of the neutrino mass matrix \cite{Chun:2003ej}.

The current search strategy is in fact sensitive to any 
{\sl singly produced} $H^{\pm\pm}$, i.e. signal candidates are
events with {\sl one pair} of same sign leptons reconstructing to 
$m_{H^{\pm\pm}}$.
This requirement is sufficient to reduce the SM background 
to negligible proportions.
Hence the search potential of the Tevatron merely 
depends on the signal efficiencies for the signal
(currently $\approx 34\%,34\%,18\%$ for $\mu\mu,ee,e\mu$) 
and the integrated luminosity. With these 
relatively high efficiencies and an expected
${\cal L}=4-8 fb^{-1}$ by the year 2009, discovery with $> 5$ events 
will be possible for $\sigma_{H^{++}H^{--}}$ of a few fb, 
which corresponds to a mass reach $m_{H^{\pm\pm}}<200$ GeV. 

Although single ${H^{\pm\pm}}$ production processes such as 
$p\overline p\to W^\pm\to
W^\mp H^{\pm\pm}$ can be neglected
\footnote{Single production of a right-handed triplet
via $q'\overline q \to W^\pm_R\to
W^\mp_R H^{\pm\pm}$ \cite{Maalampi:2002vx}
and $W^\pm_RW^\pm_R$ fusion \cite{Huitu} can be sizeable at the LHC.} 
due to the strong triplet 
VEV suppression, the mechanism $p\overline p\to W^*\to H^{\pm\pm}H^{\mp}$ 
is potentially sizeable.
This latter process proceeds via a gauge coupling constant 
and is not suppressed by any small factor. The LO partonic
cross-section is as follows:
\begin{equation}
\sigma_{LO}(q'\overline q\to H^{++}H^{-})=
{\pi\alpha^2 \over 144s_W^4 Q^2}
C_T^2p_W^2\beta_2^3
\label{singleH++}
\end{equation}
Here $C_T$ arises from the $H^{\pm\pm} H^\mp W^\mp$ vertex
and $C_T=2$ for $I=1$,$Y=2$ triplet fields 
(the doublet component of $H^\pm$ 
is negligible);
$\beta_2=\sqrt{(1-(m_{H^\pm}+m_{H^{\pm\pm}})^2/Q^2)
(1-(m_{H^\pm}-m_{H^{\pm\pm}})^2/Q^2)}$ and 
$p_W=Q^2/(Q^2-M_W^2)$.
For simplicity, we take the same $K=1.3$ as for 
$\sigma(q\overline q\to H^{++}H^{--})$ at the Tevatron
and $K=1.25$ at the LHC.
Explicit calculations for the $K$ factor for the process 
$\sigma(q'\overline q\to H^{\pm}A^0)$ in the MSSM \cite{Cao:2003tr}
(which shares the same $K$ factor as $q'\overline q\to H^{++}H^{-}$) 
give $K\approx 1.2$. In this paper we will study
in detail the magnitude and relative importance of 
$\sigma(q'\overline q\to H^{\pm\pm}H^{\mp})$. Although we
work in the HTM, our numerical analysis is relevant for 
other models which possess a $I=1$,$Y=2$ Higgs triplet
(e.g. L-R symmetric models and Little Higgs Models).

A previous quantitative study of this mechanism can be found
in \cite{Dion:1998pw}. Cross-sections were given at 
both LHC and Tevatron energies for $m_{H^{\pm\pm}}> 200$ GeV
with the simplifying assumption $m_{H^{\pm\pm}}=m_{H^{\pm}}$.
It was shown that $\sigma(q'\overline q\to H^{\pm\pm}H^{\mp})$ 
can of comparable size to $\sigma(q\overline q\to H^{++}H^{--}$).
\vspace{5mm}
\begin{figure}[h]
\begin{center}
\includegraphics[width=6.5cm,angle=-90]{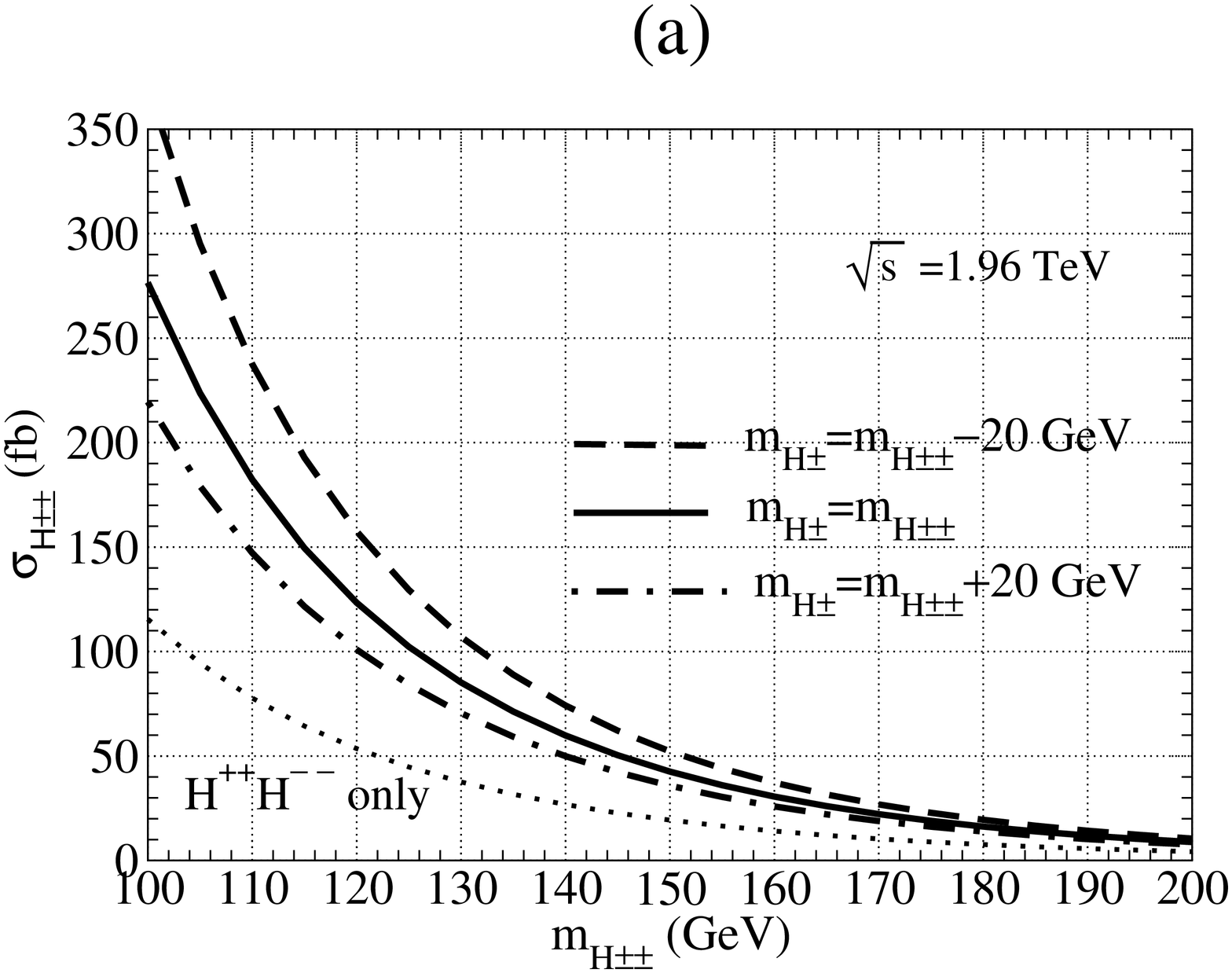} 
 \includegraphics[width=6.5cm,angle=-90]{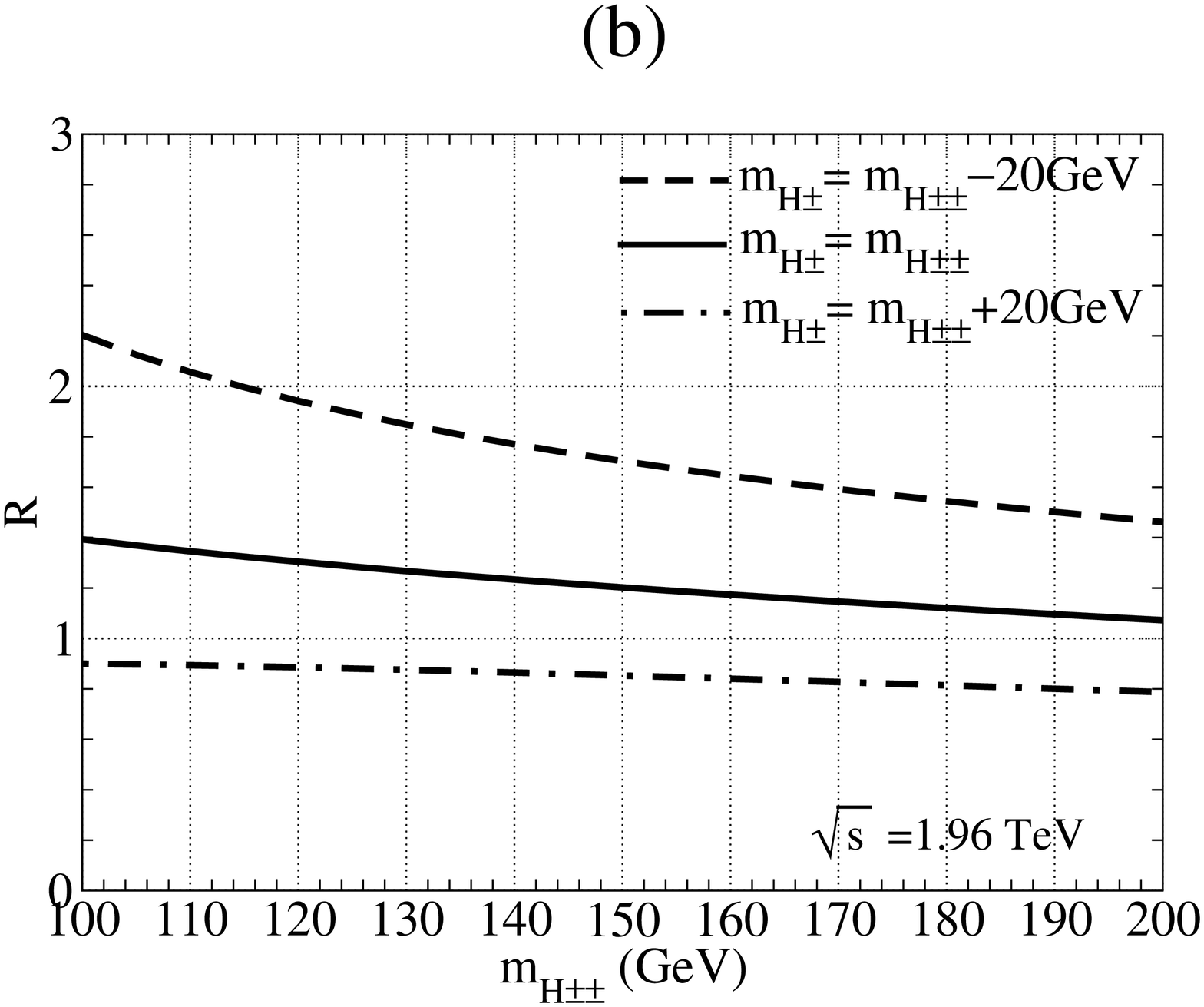}
\end{center}
\caption{%
(a)  Single production cross-section
of $H^{\pm\pm}$ ($\sigma_{H^{\pm\pm}}$)
at the Tevatron
as a function of $m_{H^{\pm\pm}}$ for different values of
$m_{H^\pm}$.
(b) Ratio $R$ as a function of $m_{H^{\pm\pm}}$.
We use CTEQ6L1 parton distribution functions.
}
\label{fig1}
\end{figure}

In this paper we first generalize the work of 
\cite{Dion:1998pw} as follows: 
\begin{itemize}

\item [{(i)}]
In our discussion at the Tevatron
we consider masses in the range 100 GeV $< m_{H^{\pm\pm}} < 200$ GeV
which will be probed during Run II, 
and allow mild mass splittings $|m_{H^{\pm\pm}}-m_{H^\pm}|\le 20$ GeV. 

\item[{(ii)}] 
In our discussion at the LHC we consider 
larger mass splittings $|m_{H^{\pm\pm}}-m_{H^\pm}|\le 80$ GeV.

\item[{(iii)}] For both the Tevatron and LHC we
study in detail the relative magnitude of
$\sigma(q'\overline q\to H^{\pm\pm}H^{\mp})$ 
and $\sigma(q\overline q\to H^{++}H^{--}$).

\end{itemize}

Moreover, motivated by the fact that the currently employed Tevatron
search strategy is sensitive to {\sl single production} of 
$H^{\pm\pm}$,
we advocate the use of the inclusive single production cross-section 
($\sigma_{H^{\pm\pm}}$) when comparing
the experimentally excluded region with the 
theoretical cross-section. This leads to a strengthening of 
the mass bound for $m_{H^{\pm\pm}}$ which
now carries a dependence on $m_{H^{\pm}}$.
We introduce the single production cross-section as follows:
\begin{equation}
\sigma_{H^{\pm\pm}}=\sigma(p\overline p,pp\to H^{++}H^{--})+
\sigma(p\overline p,pp\to H^{++} H^-)+\sigma(p\overline p,pp \to 
H^{--} H^+)
\end{equation}
At the Tevatron $\sigma(p\overline p\to H^{++} H^-)
=\sigma(p\overline p\to 
H^{--} H^+)$ while at the LHC 
$\sigma(pp\to H^{++} H^-)> \sigma(pp\to 
H^{--} H^+)$.
If a signal for $H^{\pm\pm}$ were found in the 2 lepton channel,
subsequent searches could select signal events with 3 or 4 leptons,
in order to disentangle $q\overline q\to H^{++}H^{--}$ and
$q'\overline q\to H^{\pm\pm}H^{\mp}$.
In our numerical analysis we utilize the
CTEQ6L1 parton distribution functions (pdfs) \cite{Pumplin:2002vw}.
We take the factorization scale ($Q$) as the
partonic COM energy ($\sqrt s$). 
Our results for $\sigma(q\overline q\to H^{++}H^{--})$
agree with those in \cite{Gunion:1996pq},\cite{Muhlleitner:2003me}.
Our results for $\sigma(q'\overline q\to H^{\pm\pm}H^{\mp})$
agree with those in \cite{Dion:1998pw}
(and taking $C_T=1$ agree 
with $\sigma(q'\overline q\to H^\pm A^0)$ 
in the 2HDM/MSSM \cite{Cao:2003tr}). 
The above cross-sections evaluated with MRST02 pdfs \cite{Martin:2001es}
agree with those evaluated with CTEQ6L1 to within $10\%\to 15\%$.

In Fig.\ \ref{fig1} (a) we plot $\sigma_{H^{\pm\pm}}$ as a function of 
$m_{H^{\pm\pm}}$ at the Tevatron
for three different values of $m_{H^\pm}$. We take $K=1.3$.
The current
excluded regions from the $e^\pm e^\pm,e^\pm \mu^\pm,\mu^\pm\mu^\pm$ 
searches 
correspond to the area above horizontal lines at roughly
40, 70, 35 fb respectively. The present mass limits 
for $m_{H^{\pm\pm}}$ are
where the curve for $H^{++}H^{--}$ intersects with the
above horizontal lines, and read as $133,115,136$ GeV respectively
for BR$(H^{\pm\pm}\to l^\pm_i l^\pm_j)=100\%$ .
With the inclusion of the $H^{\pm\pm}H^\mp$ channel, these
mass limits increase to 150, 130, 150 for 
$m_{H^\pm}=m_{H^{\pm\pm}}+20$ GeV, 
strengthening to 160,140,160 for $m_{H^\pm}=m_{H^{\pm\pm}}-20$ GeV. 
Clearly the search potential of 
the Tevatron (i.e. the mass limit on $m_{H^{\pm\pm}}$) 
increases significantly
when one includes the contribution to $\sigma_{H^{\pm\pm}}$
from $p\overline p\to H^{\pm\pm} H^\mp$. 
Note that the above mass limits strictly apply
to the case when $H^{\pm\pm}$ decays leptonically, and with
BR=$100\%$ in a given channel. However, if $h_{ij}$ are to 
provide the currently favoured form of the 
neutrino mass matrix then BR$(H^{\pm\pm}\to l^\pm_il^\pm_j)
< 100\%$ in a given channel. Moreover, if  
$m_{H^{\pm\pm}} > m_{H^\pm}$ then the decay channel 
$H^{\pm\pm}\to H^\pm W^*$ would be open. As shown in 
\cite{Chakrabarti:1998qy}, this decay can be sizeable 
and thus reduces BR$(H^{\pm\pm}\to l^\pm_i l^\pm_j$). 
We will return to these issues in Section 5.

In Fig.\ \ref{fig1} (b) we plot the ratio of
cross-sections $R$ at the Tevatron as
a function of $m_{H^{\pm\pm}}$, where $R$ is defined as follows:
\begin{equation}
R\equiv {\sigma(p\overline p,pp\to H^{++} H^-)
+ \sigma(p\overline p,pp \to H^{--} H^+)
\over \sigma(p\overline p,pp\to H^{++}H^{--})}
\end{equation}
The $m_{H^{\pm\pm}}$ dependence arises from the 
phase space functions $\beta_1$ and $\beta_2$ in 
Eqns.\ref{pairH++} and \ref{singleH++}.
As can be seen, $0.8 < R < 2.2$ and thus 
$q'\overline q \to H^{\pm\pm} H^\mp$ contributes significantly
to $\sigma_{H^{\pm\pm}}$.

\begin{figure}[h]
\begin{center}
\includegraphics[width=6.5cm,angle=-90]{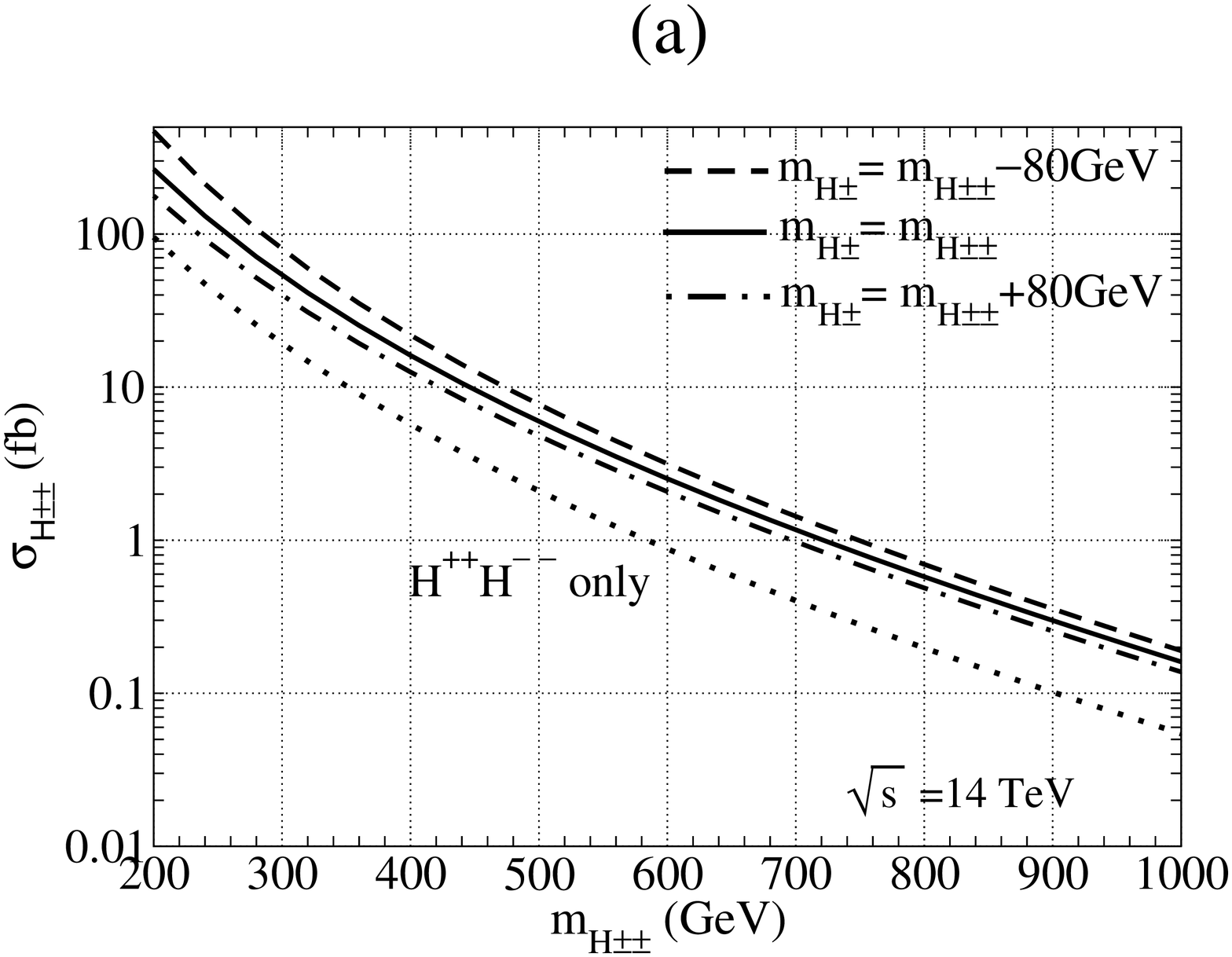}
\includegraphics[width=6.5cm,angle=-90]{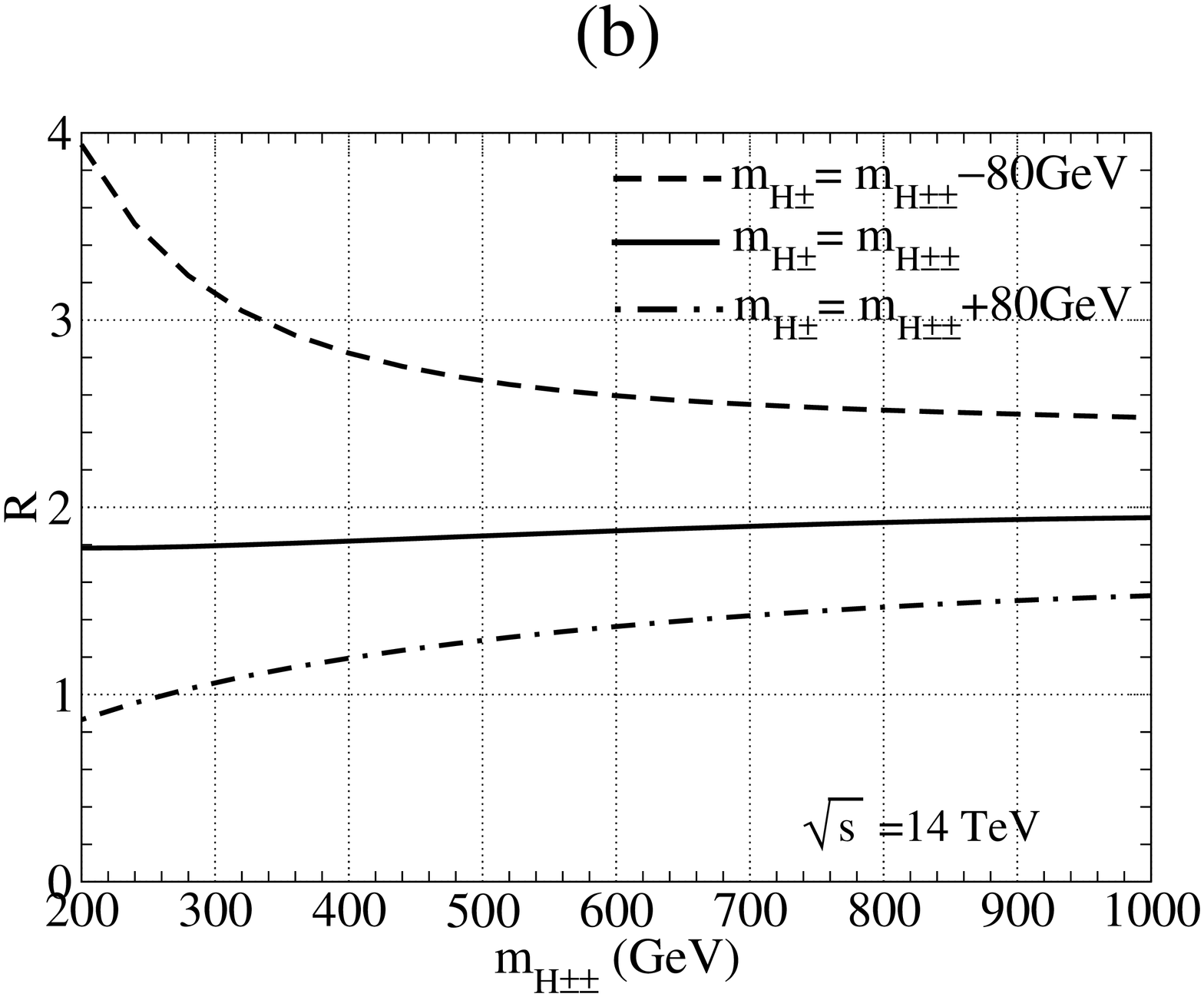}
\end{center}
\caption{%
(a)  Single production cross-section
of $H^{\pm\pm}$ ($\sigma_{H^{\pm\pm}}$)
at the LHC
as a function of $m_{H^{\pm\pm}}$ for different values of
$m_{H^\pm}$.
(b) Ratio $R$ as a function of $m_{H^{\pm\pm}}$.
We use CTEQ6L1 parton distribution functions.
}
\label{fig2}
\end{figure}

In Fig.2 we plot the analogies of Fig.1 for the LHC.
In Fig.\ \ref{fig2} (a) we plot $\sigma_{H^{\pm\pm}}$ 
for 3 values of $m_{H^{\pm\pm}}$
and for larger mass splittings ($|m_{H^{\pm\pm}}-m_{H^{\pm}}|\le 80$ 
GeV) than in Fig.\ \ref{fig2}. We take $K=1.25$.
As before, the inclusion of $q'\overline q\to H^{\pm\pm}H^{\mp}$
significantly increases the search potential 
e.g. if sensitivity to $\sigma_{H^{\pm\pm}}=1$ fb
is attained, the mass reach extends from $m_{H^{\pm\pm}}<
600$ GeV ($H^{++}H^{--}$ only)
to 750 GeV for ($m_{H^{\pm}}=m_{H^{\pm\pm}}-80$ GeV).
Recently \cite{Azuelos:2005uc} performed a simulation
of the detection prospects at the LHC for $q\overline q\to H^{++}H^{--}$ 
for the cases where 3 and 4 leptons are detected. 
With 100 fb$^{-1}$, sensitivity to $m_{H^{\pm\pm}}\lsim 800$ GeV 
(3 leptons) and 
$m_{H^{\pm\pm}}\lsim 700$ GeV (4 leptons) is expected. We are not aware of
a simulation for the case where only 2 leptons are detected.
Presumably even larger values of $m_{H^{\pm\pm}}$ ($\gsim 800$ GeV) 
could be probed.
In  Fig.\ \ref{fig2} (b) we plot $R$ as a function of $m_{H^{\pm\pm}}$.
One can see that $R>1$ for the upper two curves for all
$m_{H^{\pm\pm}}$, while for the lower curve $R>1$ for
$m_{H^{\pm\pm}}>260$ GeV. 
Note that the dependence of $R$ on $m_{H^{\pm\pm}}$
differs from that observed in Fig.\ \ref{fig1} (b),
which can be attributed to the
different parton luminosity functions at the Tevatron and LHC.

\vspace{5mm}

\section{Neutrino mass hierarchy and the decay $H^{\pm\pm}\to H^\pm W^*$}

The current experimental searches assume 
that the sole decay mode of $H^{\pm\pm}$ is 
$H^{\pm\pm}\to l^\pm_il^\pm_j$ mediated by the 
arbitrary Yukawa couplings $h_{ij}$.
The decay rate for $H^{\pm\pm}\to l^\pm_il^\pm_j$ is given by:
\begin{equation}
\Gamma(H^{\pm\pm}\to l_i^\pm l_j^\pm)=
S{m_{H^{\pm\pm}}\over 8\pi}|h_{ij}|^2
\label{Hlldecay}
\end{equation}
where $S=1(2)$ for $i=j$ ($i\ne j$). 
Clearly $\Gamma(H^{\pm\pm}\to l_i^\pm l_j^\pm)$ 
depends crucially on the {\sl absolute} value of the
$h_{ij}$, although the leptonic BRs are determined by the 
{\sl relative} values.
In this section we consider the impact of the decay mode
$H^{\pm\pm}\to H^\pm W^*$ on the BRs of the leptonic channels.
It has been known for some time that BR($H^{\pm\pm}\to H^\pm W^*$)
is potentially sizeable and a quantitative analysis can be found
in \cite{Chakrabarti:1998qy}. The decay rate for 
$H^{\pm\pm}\to H^\pm W^*$ (summing over all
fermion states for  $W^*\to ff$ excluding the $t$ quark) is given by:
\begin{equation}
\Gamma(H^{\pm\pm}\to H^\pm W^*)=9G_F^2M_W^4m_{H^{\pm\pm}}C_T^2P/(16\pi^3)
\label{HHWdecay}
\end{equation}
where $P$ is the phase space term (which we calculate by 
numerical integration) and $C_T(=2)$ is from the coupling 
$H^{\pm\pm}H^\pm W$. $P$ depends on the 
mass difference $\Delta m$ defined by
$\Delta m= m_{H^{\pm\pm}}-m_{H^\pm}$, and $P=0$ for 
$\Delta m=0$.
If $m_{H^\pm}<m_{H^{\pm\pm}}$ this decay 
can compete with $H^{\pm\pm}\to l_i^\pm l_j^\pm$ since the
phase space suppression of the virtual $W^*$ is compensated by
the gauge strength coupling \cite{Gunion:1996pq}.
Ref.\cite{Chakrabarti:1998qy} showed that
$H^{\pm\pm}\to H^\pm W^*$ can dominate over 
$H^{\pm\pm}\to l_i^\pm l_j^\pm$ if $\Delta m$ is sizeable
($> 40$ GeV) and $h_{ij}$ are of order $10^{-3}$ or less.
A large BR($H^{\pm\pm}\to H^\pm W^*$) would
debilitate the $H^{\pm\pm}$ search potential in the leptonic channel.
However, as emphasized in \cite{Gunion:1998ii},
observation of $H^{\pm\pm}\to H^\pm W^*$ together with
one or more of the leptonic channels could provide
information on the absolute values of $h_{ij}$.
If only BR($H^{\pm\pm}\to l_i^\pm l_j^\pm$) are
measured then only the {\sl relative} values of the 
$h_{ij}$ can be evaluated.
The decay rate for $H^{\pm\pm}\to H^\pm W^*$ is 
theoretically calculable once 
$m_{H^\pm}$ and $m_{H^{\pm\pm}}$ are known experimentally,
and thus it can be used as a benchmark decay with which
to estimate the total width of $H^{\pm\pm}$.
It is known that the BRs of the 
leptonic channels depend on which solution to the
neutrino mass matrix is realized \cite{Chun:2003ej}.
However, a quantitative analysis of the 
impact of $H^{\pm\pm}\to H^\pm W^*$ in the 
various allowed scenarios is still lacking and will be presented below.
We are not aware of any experimental simulation of
$H^{\pm\pm}\to H^\pm W^*$. The signature would depend crucially
on the decay products of $H^\pm$, which are
are either $H^\pm\to l^\pm\nu_l$ (driven by $h_{ij}$),
or possibly $H^\pm\to H^0W^*,A^0W^*$.

We now briefly review relevant results and formulae
from neutrino physics.
The neutrino mass matrix is diagonalized by the MNS 
(Maki-Nakagawa-Sakata) matrix $V_{_{\rm MNS}}$ \cite{MNS}.
Using Eq.(\ref{nu_mass}) one can write the couplings
$h_{ij}$ as follows:
\begin{equation}
h_{ij}=\frac{1}{\sqrt{2}v_\Delta}V_{_{\rm MNS}}diag(m_1,m_2,m_3)
V_{_{\rm MNS}}^T
\label{hij}
\end{equation}
Here we take the basis in which the unitary matrix responsible 
for diagonalizing the
charged-lepton mass matrix is a unit matrix.
The MNS matrix in the standard parametrization
is as follows:
\begin{equation}
V_{_{\rm MNS}}^{} =
\bmaT
c_1c_3                        & s_1c_3                  & s_3e^{-i\delta} \\
-s_1c_2-c_1s_2s_3e^{i\delta}  & c_1c_2-s_1s_2s_3e^{i\delta}  & s_2c_3 \\
s_1s_2-c_1c_2s_3e^{i\delta}   & -c_1s_2-s_1c_2s_3e^{i\delta} & c_2c_3  
\emaT
\bmaT
1 & 0 & 0 \\
0 & e^{i \varphi_1^{}/2} & 0 \\
0 & 0 & e^{i \varphi_2^{}/2} 
\ema
\,,
\end{equation} 
where $s_i\equiv\sin\theta_i$ and $c_i\equiv \cos\theta_i$, $\delta$ is
the Dirac phase and $\varphi_1$ and $\varphi_2$ are the Majorana phases.

Neutrino oscillation experiments involving solar \cite{SNO}, atmospheric 
\cite{atm} and reactor neutrinos \cite{RCT}
are sensitive to the mass-squared 
differences and the mixing angles. 
and give the following preferred values:
\begin{eqnarray}
\Delta m^2_{12} \equiv m^2_2 -m^2_1
\simeq 8.0\times 10^{-5} {\rm eV}^2 \,,~~
|\Delta m^2_{13}|\equiv |m^2_3 -m^2_1|
\simeq 2.1\times 10^{-3} {\rm eV}^2\,, \\
\sin^22\theta_{1}\simeq 0.8 \,,~~~~ \sin^22\theta_{2}\simeq 1 \,,~~~~
\sin^22\theta_{3}\lsim 0.16\,.~~~~~~~~~~~~
\label{obs_para}
\end{eqnarray}
Since the sign of $\Delta m_{13}^2$ and the mass of the lightest 
neutrino are both undetermined at present, 
distinct neutrino mass hierarchy patterns are classified
as follows: 
{\it Normal hierarchy} (NH) ($m_1 < m_2 \ll m_3$), 
{\it Inverted hierarchy} (IH) ($m_2 >  m_1 \gg m_3$),  
{\it Quasi-degenerate} (DG) ($m_1 \sim m_2 \sim m_3 \gg 
\sqrt{|\Delta m^2_{13}|}$).
From Eq.(\ref{nu_mass}) and Eq.(\ref{hij})
it can be shown that: 
\begin{equation}
\sum_{i,j} h_{ij}^2v^2_{\Delta}\propto \sum_i m^2_i\,,
\label{nu_mass1}
\end{equation}
Hence the total leptonic decay width depends on the absolute mass 
of the neutrinos, and the value of $\sum_i m^2_i$ depends on
which solution to the neutrino mass matrix (NH,IH,DG)
is realized. The minimum value of $\sum_i m^2_i$ is 
$|\Delta m^2_{13}|$
while the maximum is given by the cosmological constraint.

\begin{figure}[ht]
\begin{center}
\includegraphics[width=4.3cm,angle=-90]{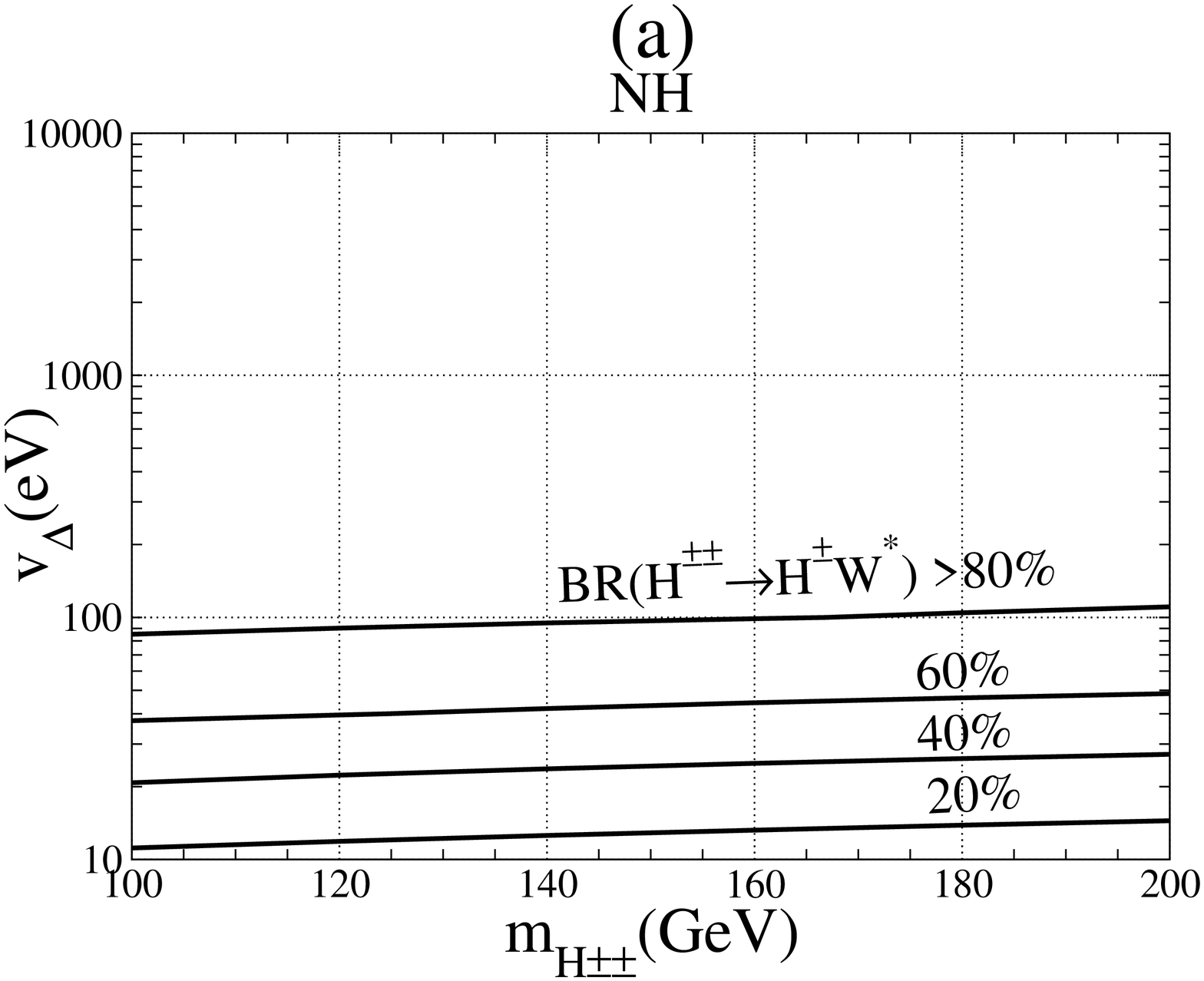}
\includegraphics[width=4.3cm,angle=-90]{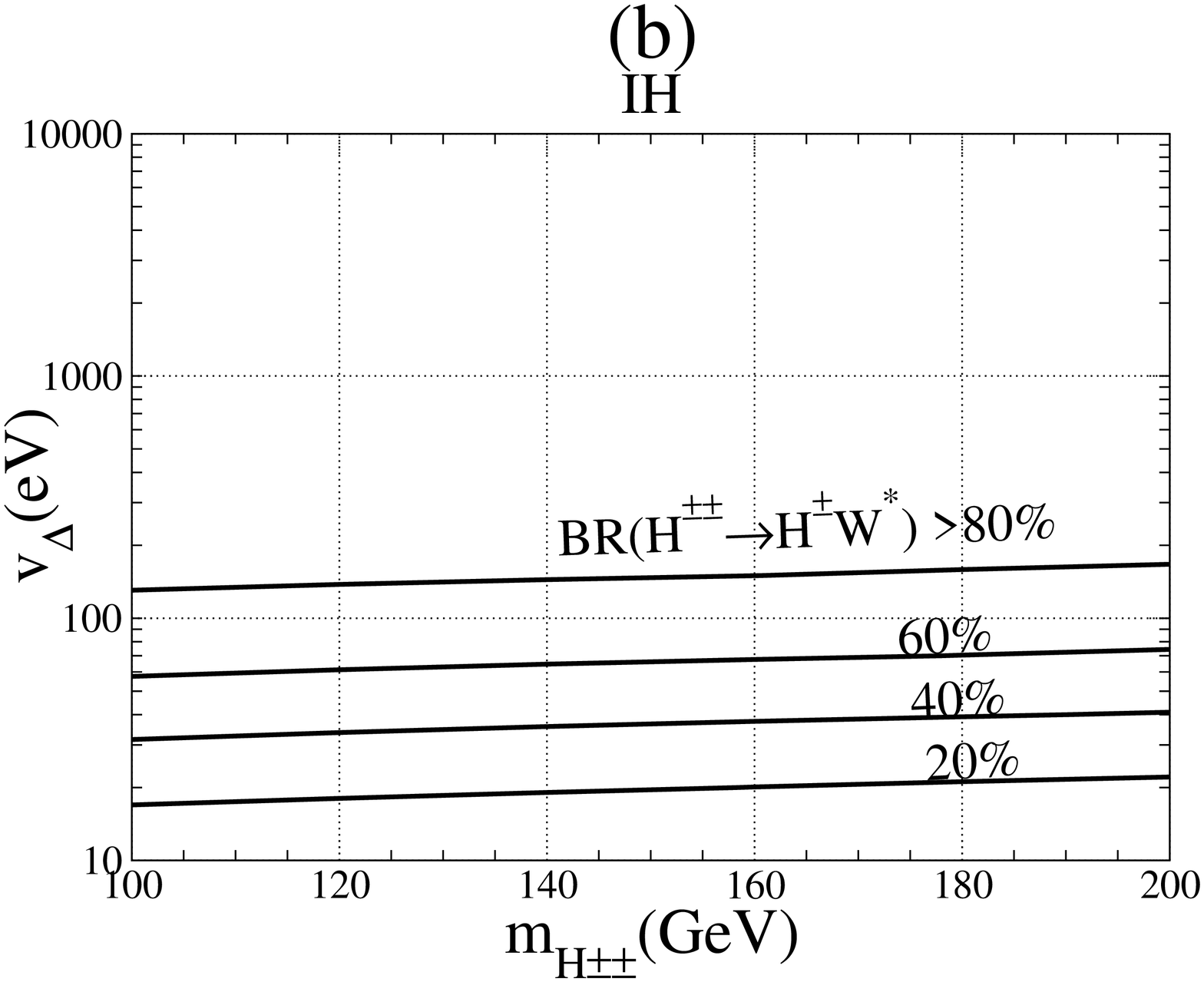}
\includegraphics[width=4.3cm,angle=-90]{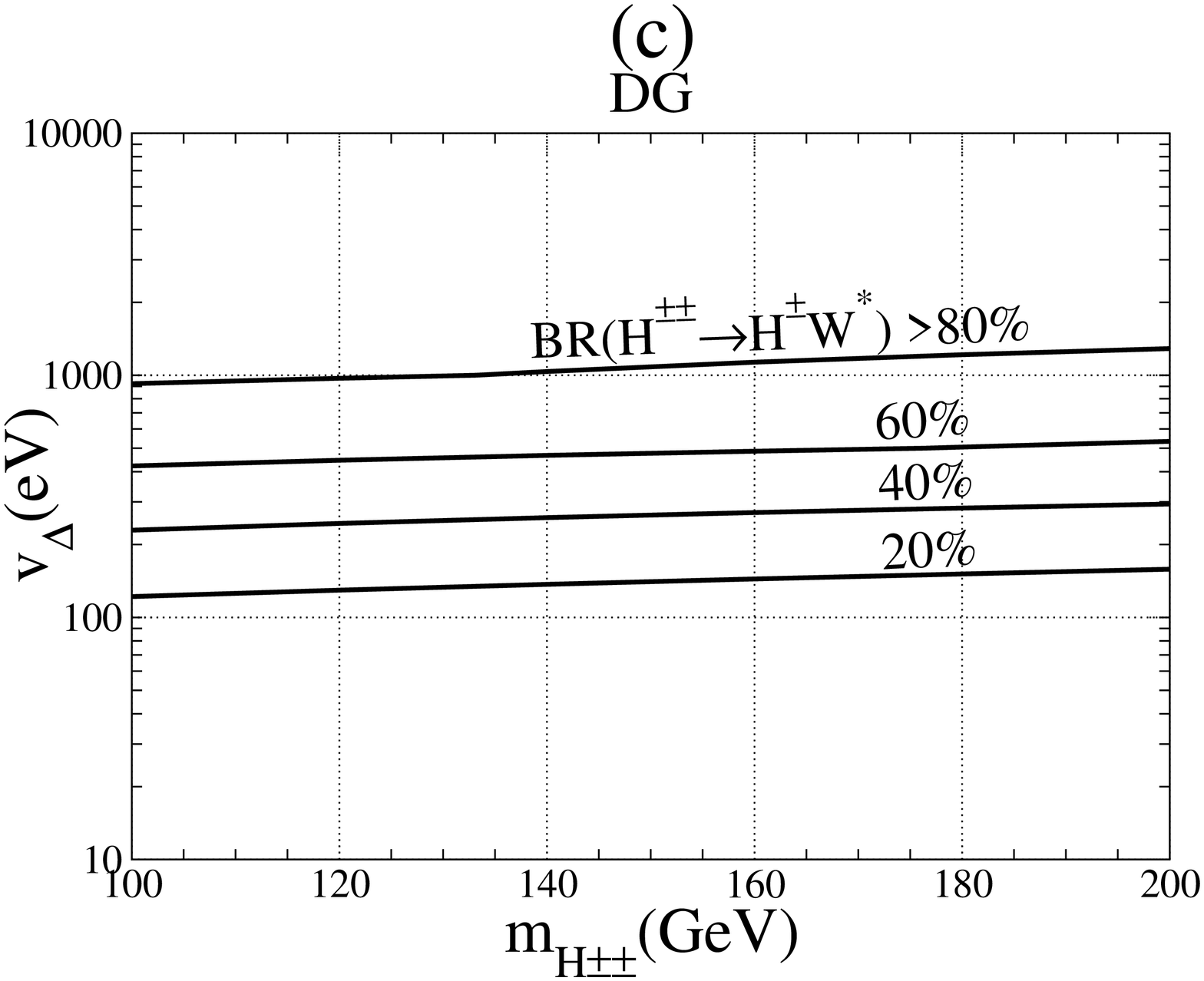}
\caption{%
Contours of BR$(H^{\pm\pm}\to H^\pm W^*)$ 
in the plane ($m_{H^{\pm\pm}}, v_{\Delta}$), 
for NH (a), IH (b) and DG (c). We take 
$m_{H^\pm}=m_{H^{\pm\pm}}-20$ GeV.}
\label{fig3}
\end{center}
\end{figure}

In Fig.\ref{fig3} we show contours of BR$(H^{\pm\pm}\to H^\pm W^*)$ 
in the plane ($m_{H^{\pm\pm}}, v_{\Delta}$), for three different
solutions to the neutrino mass matrix.
We assume that $m_{1(3)}=0$ for NH (IH) and $m_1=0.2$ eV for DG.
We take $m_{H^\pm}=m_{H^{\pm\pm}}-20$ GeV.
From Eq.\ref{hij}, all $h_{ij}$ are determined once 
$v_\Delta$ is specified. In order to comply
with current experimental upper limits on
LFV decays of $\mu^\pm$ and $\tau^\pm$, one can
derive the bound $v_{\Delta}> 10$ eV 
for NH and IH, and $v_{\Delta}> 100$ eV
for DG. The stronger constraint on $v_{\Delta}$
in DG arises because
$\sum_im_i$ in DG is larger than those in NH and IH.

From Fig.\ref{fig3} it is clear that BR$(H^{\pm\pm}\to H^\pm W^*)$ 
can be sizeable, and approaches $100\%$ for larger
$v_{\Delta}$. For a fixed value of $v_{\Delta}$, 
one can see that BR$(H^{\pm\pm}\to H^\pm W^*)$ is
relatively more important in NH and IH than in DG.
This can be understood from Eq.~\ref{nu_mass1},
since DG requires heavier neutrinos (and thus larger
$h_{ij}$) which in turn reduces BR$(H^{\pm\pm}\to H^\pm W^*)$.
One can consider three distinct scenarios 
with very different magnitudes for 
BR$(H^{\pm\pm}\to H^\pm W^*)$ and BR$(H^{\pm\pm}\to l^\pm l^\pm$):

\begin{itemize}

\item[{(i)}] BR$(H^{\pm\pm}\to H^\pm W^*) \gg$ 
BR$(H^{\pm\pm}\to l^\pm l^\pm$):\\
In this case the current search strategy
(which requires $H^{\pm\pm}\to l^\pm l^\pm$ decay) is ineffective.
Simulations have not been carried out for the decay
$H^{\pm\pm}\to H^\pm W^*$ although one might naively 
expect sensitivity
comparable to that for the decay $H^{\pm\pm}\to \tau^\pm\tau^\pm$,
as suggested in \cite{Gunion:1996pq},\cite{Gunion:1998ii}.

\item[{(ii)}] BR$(H^{\pm\pm}\to H^\pm W^*)\approx$
BR$(H^{\pm\pm}\to l^\pm l^\pm$):\\
The search for
$H^{\pm\pm}\to l^\pm l^\pm$ would be effective and
$H^{\pm\pm}$ could be discovered in one or more leptonic channels.
If $H^{\pm\pm}\to H^\pm W^*$ is also observed
then information on the absolute value
of $h_{ij}$ might be possible:
Using Eqs.\ref{Hlldecay} and \ref{HHWdecay},
the ratio of leptonic events ($N_{l_il_j}$) 
to $H^\pm W^*$ events ($N_{H^\pm W^*}$) 
is given as follows:
\begin{equation}
{N_{l_il_j}\over N_{H^\pm W^*}}\sim {h^2_{ij}\over P}
\end{equation}

Observation of the leptonic channel
provides $m_{H^{\pm\pm}}$. If $m_{H^\pm}$ can be roughly measured
then $P$ (and hence the partial width for $H^{\pm\pm}\to H^\pm W^*$) 
can be calculated. From the above equation one can obtain an 
order of magnitude estimate of $h_{ij}$.

\item[{(iii)}]BR$(H^{\pm\pm}\to H^\pm W^*) \ll $ 
BR$(H^{\pm\pm}\to l^\pm l^\pm$):\\
In this case the current search strategy
($H^{\pm\pm}\to l^\pm l^\pm$) is effective.
If BR$(H^{\pm\pm}\to l^\pm_i l^\pm_j$) are measured
then the ratios of $h_{ij}$ can be evaluated. This can be
compared with Eqn.\ref{nu_mass} in order to see which
neutrino solution is realized \cite{Chun:2003ej}.
The absolute values of $h_{ij}$ cannot be measured 
unless a LFV decay of $\mu$ and/or $\tau$ is observed.

\end{itemize}

\section{Tevatron search potential in HTM}

We now study the search potential of the Tevatron for the
generalized case in the HTM where 
$p\overline p\to H^{\pm\pm}H^{\mp}$ is included,
BR$(H^{\pm\pm}\to H^\pm W^*)\ne 0\%$ 
and $h_{ij}$ are required to reproduce a 
phenomenologically acceptable neutrino mass matrix.
We relax the assumptions for the Majorana phases and take 
$\varphi_1,\varphi_2=0$ or $\pi$, which leads the 7 distinct solutions:
\begin{center}
\begin{tabular}{lllll}
NH: & $m_1 < m_2 \ll m_3$,&~~~~ &&  \\
IH1:& $m_2 > m_1 \gg m_3$,& &IH2: & $ -m_2 > m_1 \gg m_3$,  \\
DG1:& $m_1 \simeq m_2 \simeq m_3$,& &DG2: & $m_1 \simeq m_2 \simeq -m_3$, \\
DG3:& $m_1 \simeq -m_2 \simeq m_3$,& &DG4: & $m_1 \simeq -m_2 \simeq -m_3$. \\
\end{tabular}
\end{center}
In the HTM, BR$(H^{\pm\pm}\to l^\pm l^\pm)$ are predicted and different
in each of the 7 distinct solutions 
(NH,IH1,IH2,DG1$\to$DG4), and their ratios were evaluated in
\cite{Chun:2003ej}. 
Note that such predictions of BR$(H^{\pm\pm}\to l^\pm l^\pm)$
are a feature of the HTM in which
the couplings $h_{ij}$ are the sole origin of
neutrino mass. This direct correlation between 
BR$(H^{\pm\pm}\to l^\pm l^\pm)$ and the neutrino mass matrix
may not extend to $H^{\pm\pm}$ of other models in which 
neutrinos can acquire mass by other means
e.g. the seesaw mechanism in L-R models or
by a combination of mechanisms which may or may not include the
$h_{ij}$ couplings \cite{Han:2005nk},\cite{Diaz:1998zg}.
In contrast, the production process
 $\sigma(p\overline p\to H^{\pm\pm}H^{\mp}$)
is certainly relevant in any model with $Y=2$ triplets.

\begin{figure}[h]
\begin{center}
\includegraphics[width=6.5cm,angle=-90]{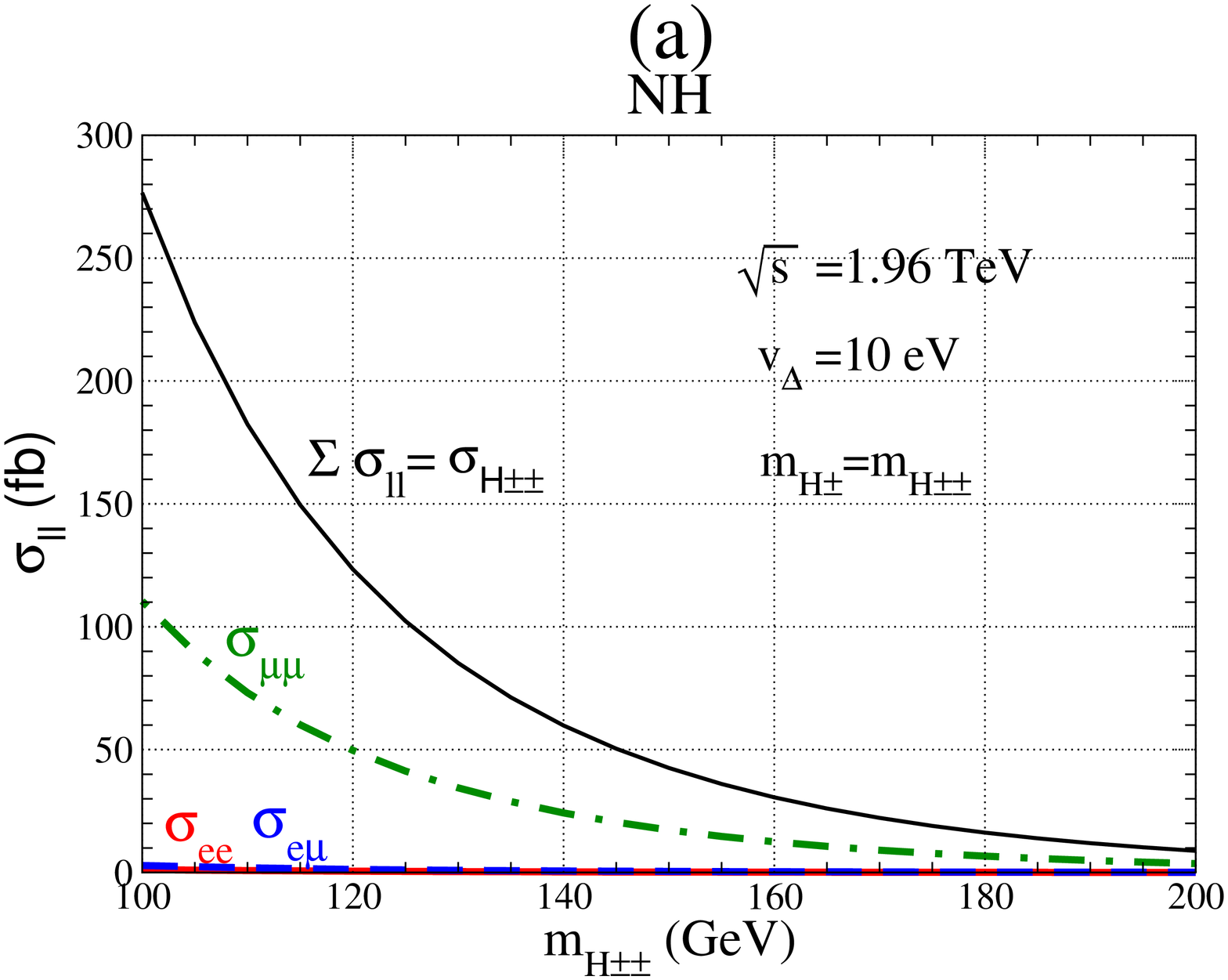}
\includegraphics[width=6.5cm,angle=-90]{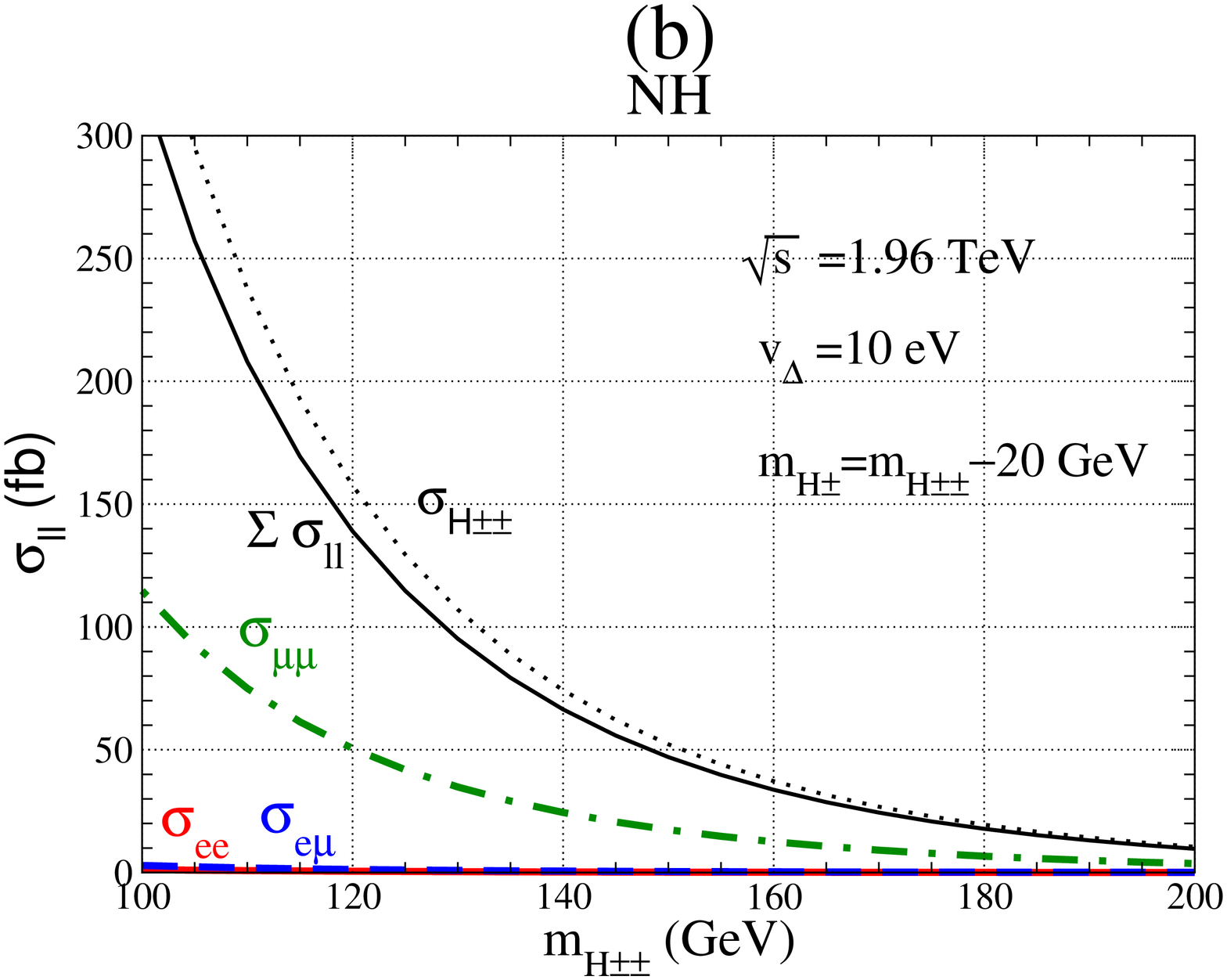}
\end{center}
\caption{%
$\sigma_{ll}$ as a function of $m_{H^{\pm\pm}}$
for NH with (a) $m_{H^{\pm}}=m_{H^{\pm\pm}}$ 
and (b) $m_{H^{\pm}}=m_{H^{\pm\pm}}-20$ GeV .}
\label{fig4}
\end{figure}

In Figs.\ref{fig4}$\to$ \ref{fig6}
we plot $\sigma_{ll}$ as a function of $m_{H^{\pm\pm}}$,
where $\sigma_{ll}$ is the total leptonic ($l=e,\mu,\tau$)
cross-section defined as:
\begin{equation}
\sigma_{ll}=\sigma(p\overline p\to H^{++}H^{--})\times
B_{ll}(2-B_{ll})+2\sigma(p\overline p\to H^{++}H^{-})\times B_{ll}
\label{crossLL}
\end{equation}
The contribution to $\sigma_{ll}$ from 
$\sigma(p\overline p\to H^{++}H^{--})$ falls more slowly with 
decreasing $B_{ll}$ 
since signal candidates are events with at least 2 leptons.
Eq.(\ref{crossLL}) simplifies to Eq.\ref{crossBR}
in the limit where $\sigma(p\overline p\to H^{\pm\pm}H^{\mp})=0$
and $B_{ll}=1$. 
Figs.\ref{fig4}(a) shows $\sigma_{ll}$ for the NH
with $m_{H^\pm}=m_{H^{\pm\pm}}$, which leads to $B_{ll}=1$. 
In this case $\sum\sigma_{ll}=\sigma_{H^{\pm\pm}}$.
For the other figures we take $m_{H^\pm}=m_{H^{\pm\pm}}-20$ GeV, 
which induces a sizeable (but not dominant) 
BR($H^{\pm\pm}\to H^\pm W^*)$, and hence 
$\sum\sigma_{ll}<\sigma_{H^{\pm\pm}}$.
We set $v_\Delta=10$ eV in Figs.\ref{fig4} and 5
and $v_\Delta=100$ eV in Figs.\ref{fig6}.
We only plot $\sigma_{ll}$ for $ee,e\mu,\mu\mu$
since the Tevatron has already performed searches in these channels.
Sensitivity to $\sigma_{ll}$ of a few fb will be possible with
the anticipated integrated luminosities of $4-8$ fb$^{-1}$.
There are plans to search for the 3 leptonic decays involving $\tau$
($e\tau,\mu\tau,\tau\tau$) although the discovery reach in 
$m_{H^{\pm\pm}}$ is expected to be inferior to that for the 
$ee,e\mu,\mu\mu$ channels.
In all figures we take $\theta_3=0^\circ$.
From the figures it is clear that 
$\sigma_{ee,e\mu,\mu\mu}$ differ considerably in each of the
7 scenarios. Optimal coverage is for cases DG1 and DG4, which
have $\sigma_{ee,\mu\mu}\ge 5$ fb and
$\sigma_{e\mu,\mu\mu}\ge 5$ fb respectively
for $m_{H^{\pm\pm}}\lsim 180$ GeV.
For NH, $\sigma_{\mu\mu}\ge 5$ fb for $m_{H^{\pm\pm}}\lsim 190$ GeV
but $\sigma_{ee}$ and $\sigma_{e\mu}$ are both unobservable. 
Taking $\theta_3$ at its largest experimentally allowed value
results in minor changes to all figures, with the 
most noticeable effect
being a significant reduction of $\sigma_{\mu\mu}$ in DG4.
Clearly the Tevatron Run II not only has 
strong search potential for $H^{\pm\pm}$, 
but is also capable of distinguishing between the various 
allowed scenarios for the neutrino mass matrix.

\vspace{5mm}

\vspace{5mm}
\begin{figure}[h]
\begin{center}
\includegraphics[width=6.5cm,angle=-90]{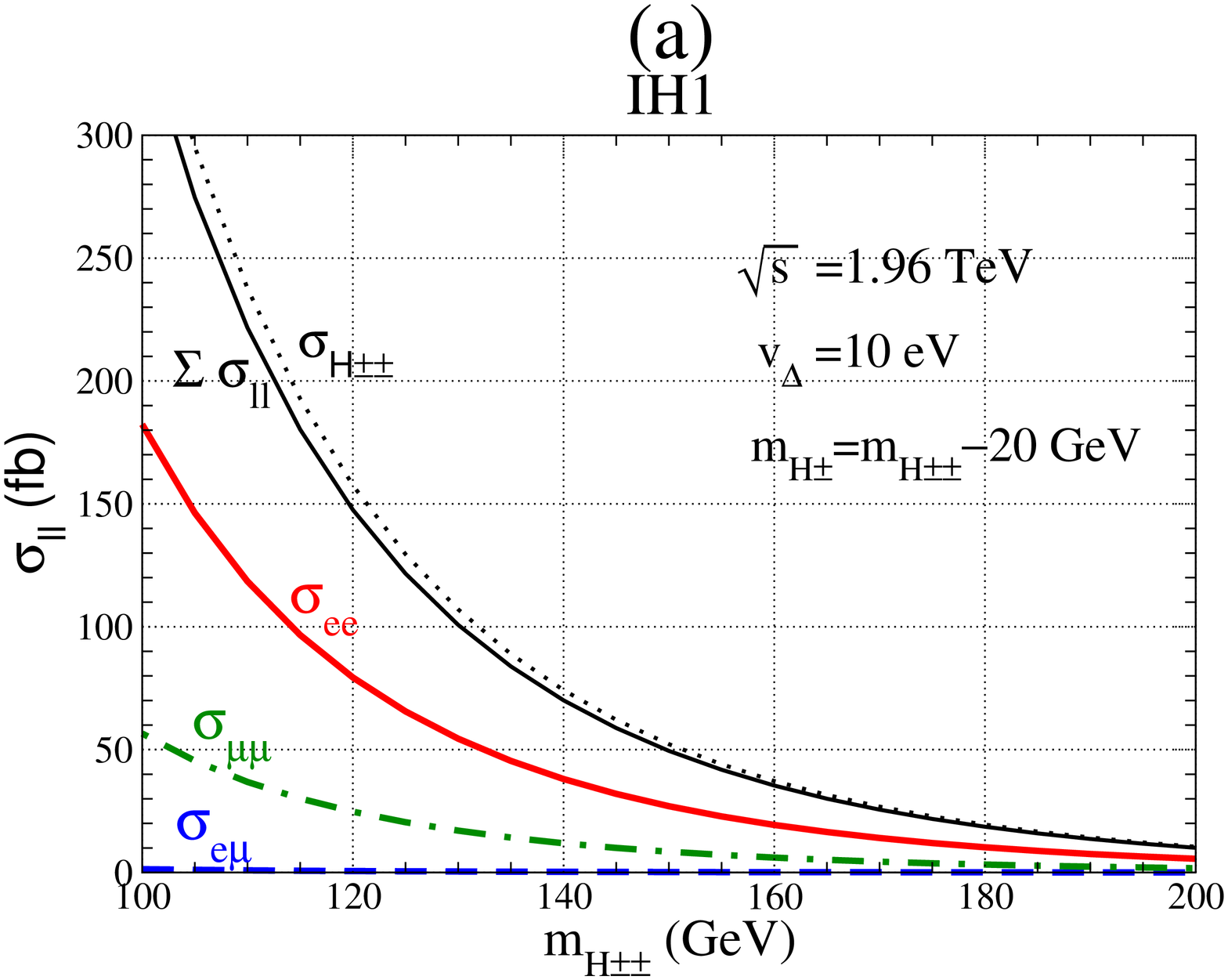}
\includegraphics[width=6.5cm,angle=-90]{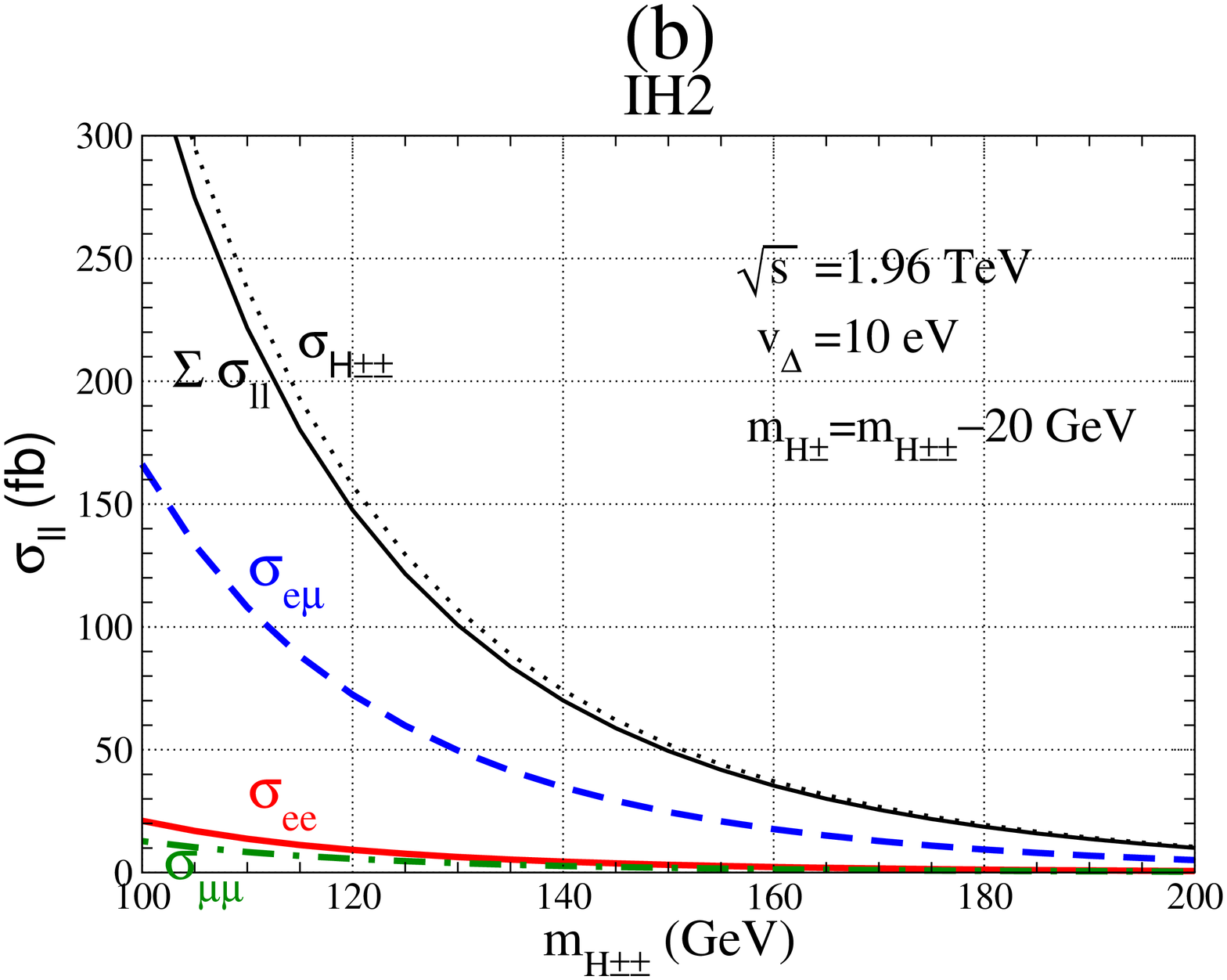}
\end{center}
\caption{%
$\sigma_{ll}$ as a function of $m_{H^{\pm\pm}}$
for (a) IH1 and (b) IH2.}
\label{fig5}
\end{figure}

\vspace{5mm}
\begin{figure}[h]
\begin{center}
\includegraphics[width=6.5cm,angle=-90]{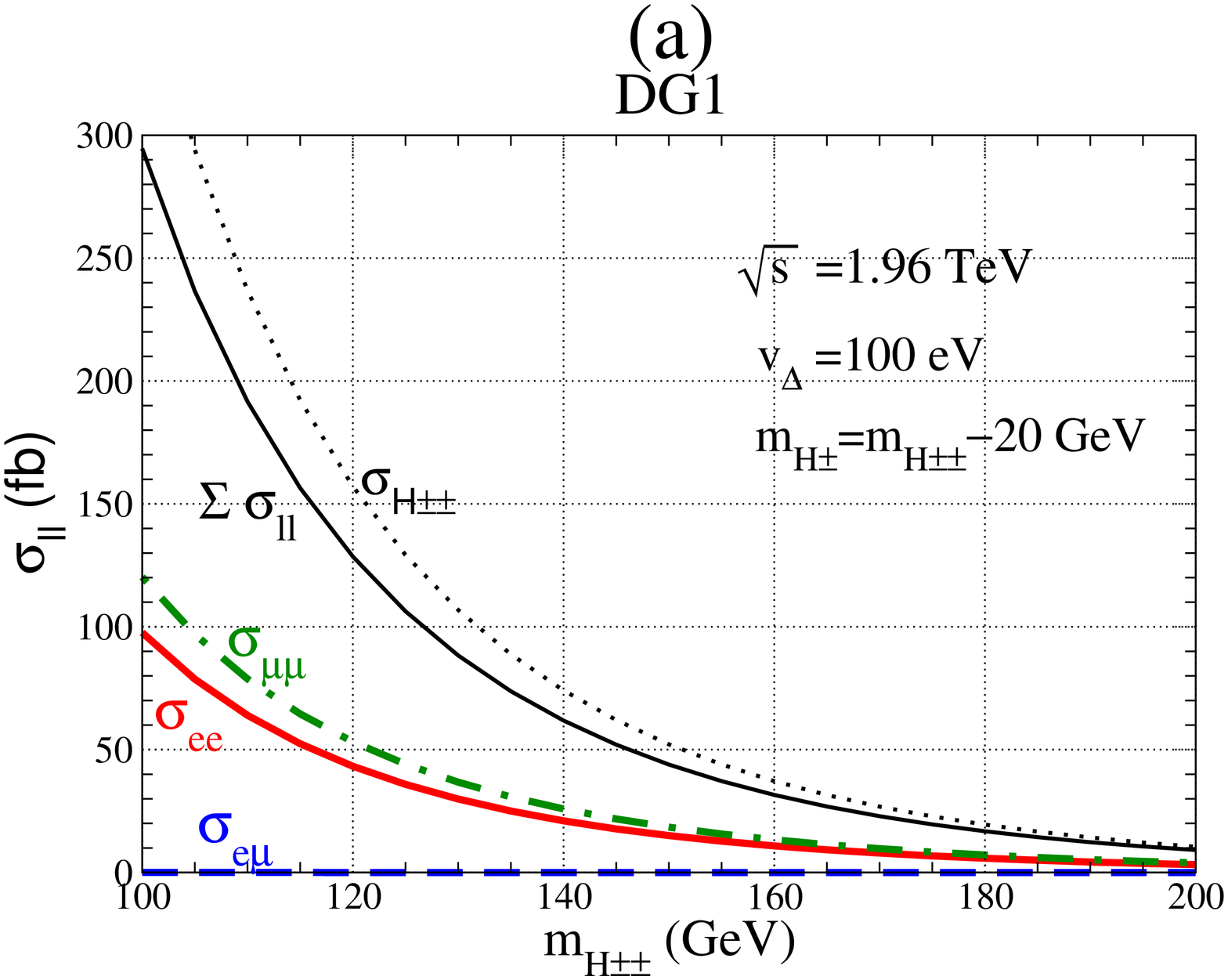}
\includegraphics[width=6.5cm,angle=-90]{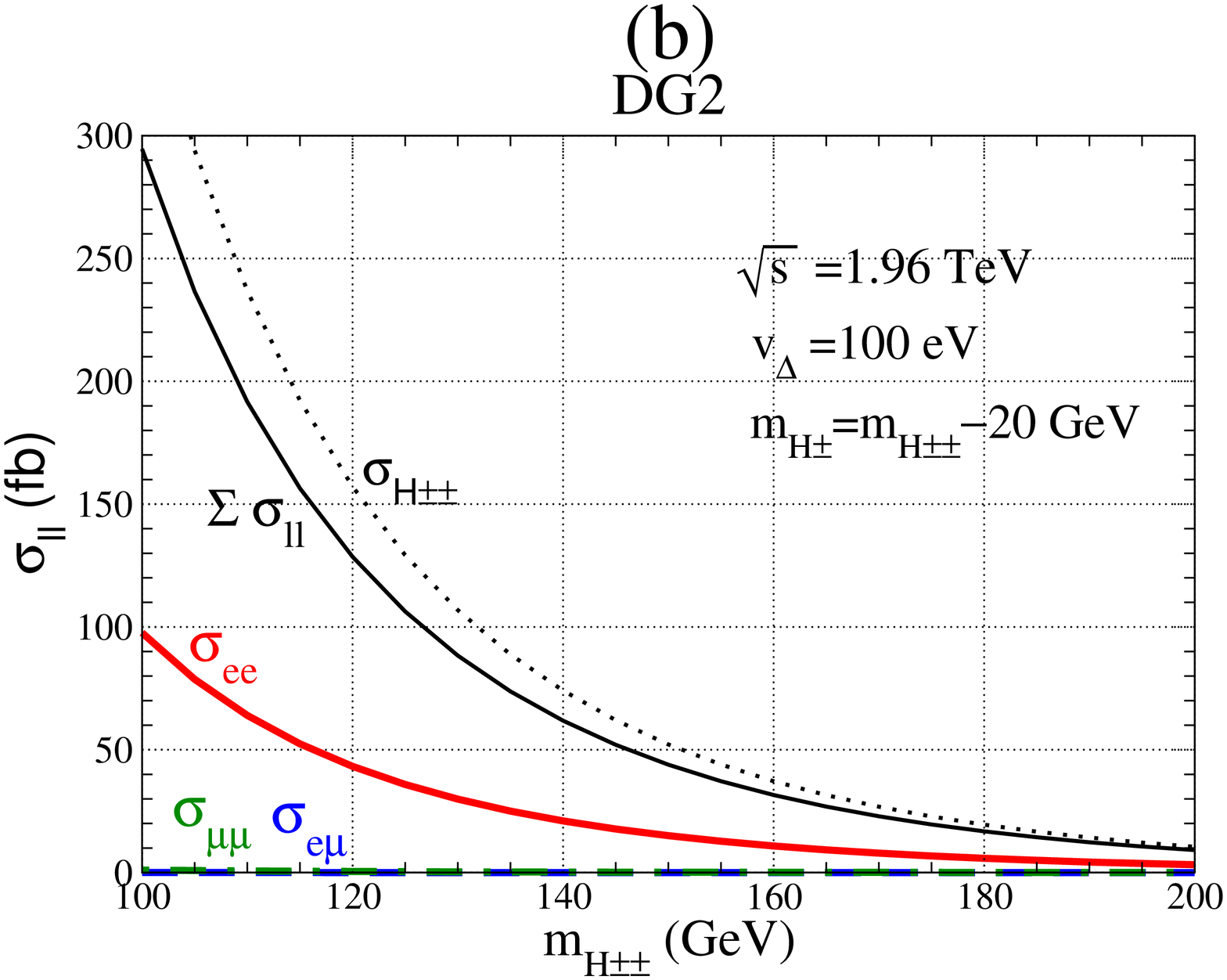} \\
\includegraphics[width=6.5cm,angle=-90]{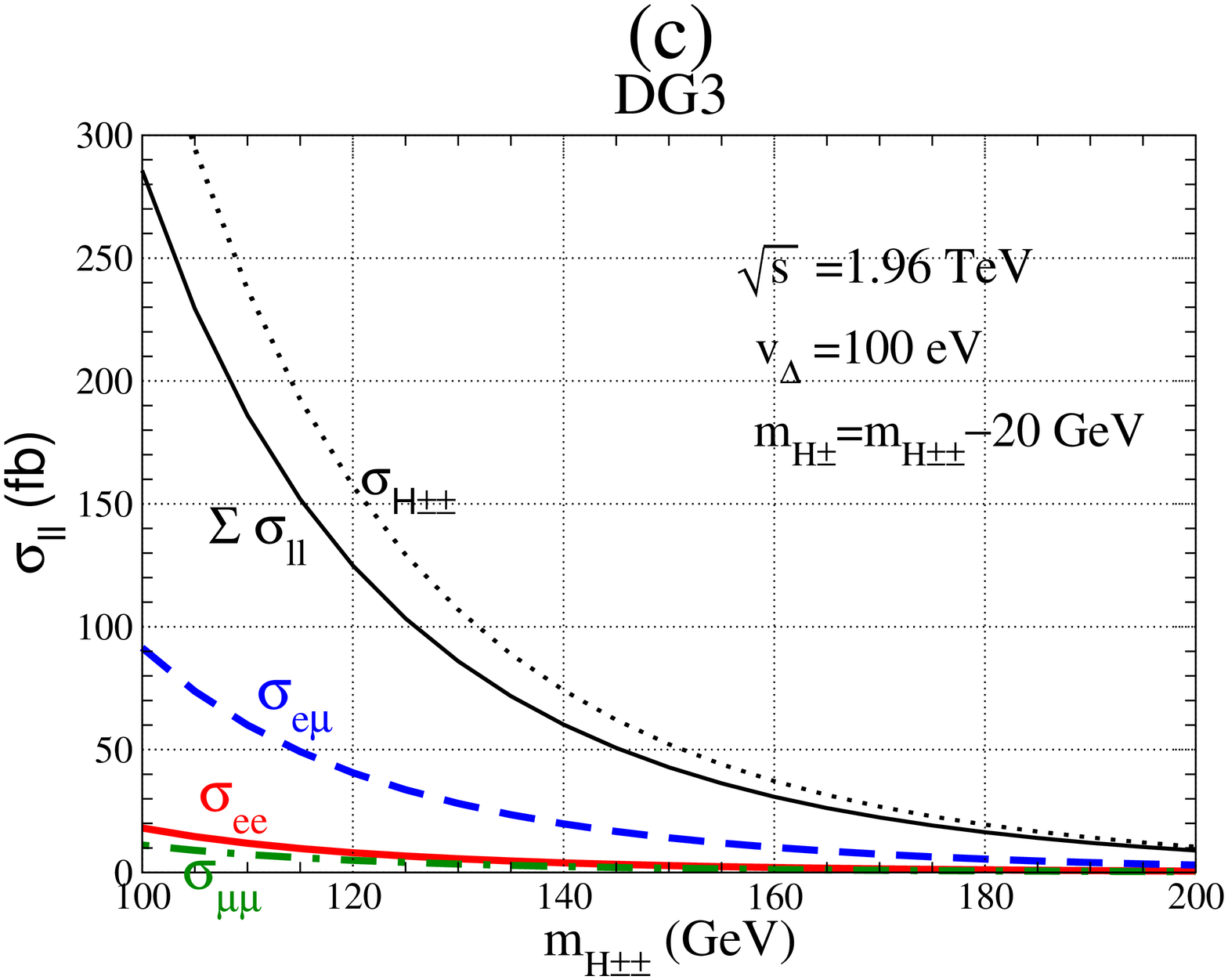}
\includegraphics[width=6.5cm,angle=-90]{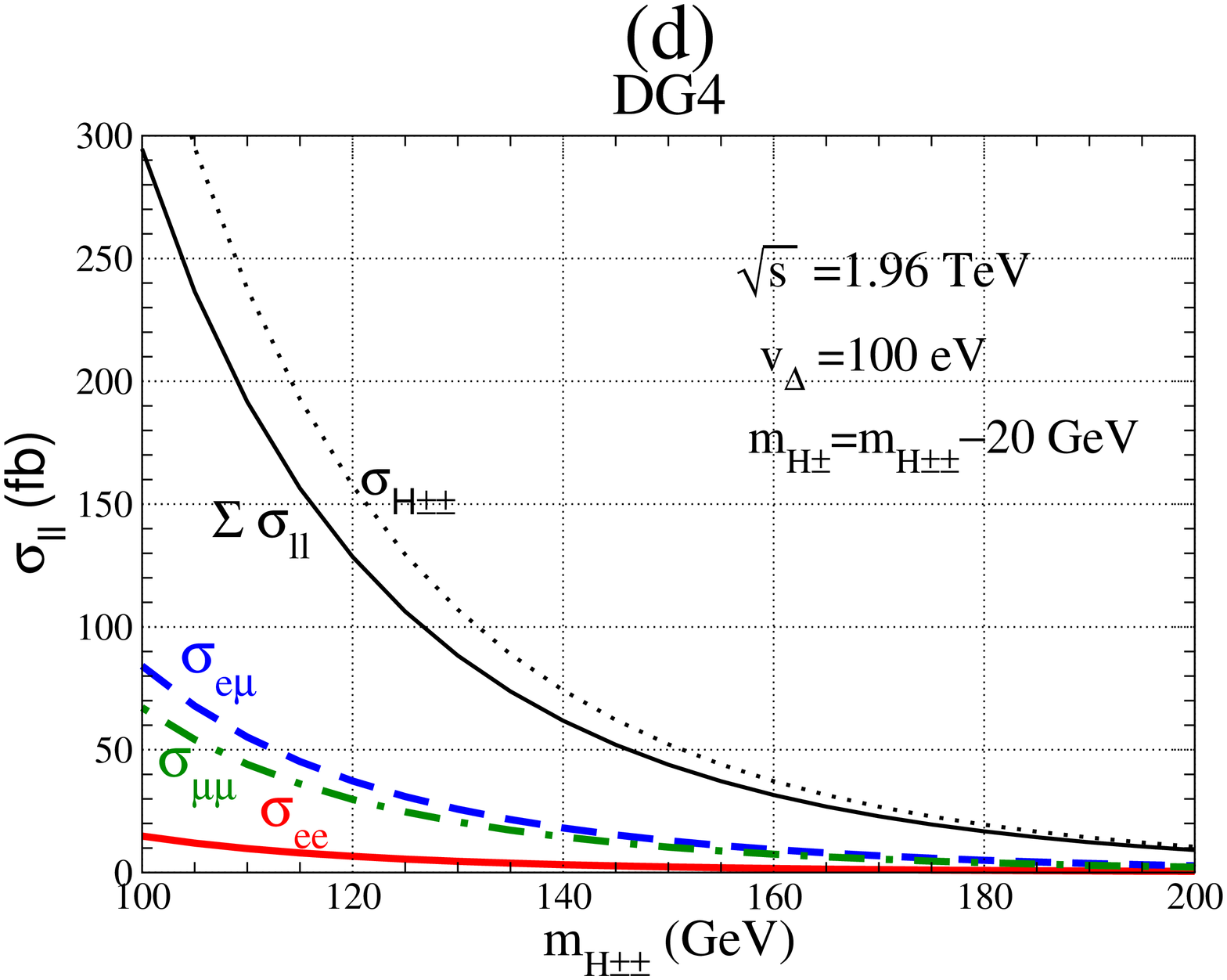}
\end{center}
\caption{%
$\sigma_{ll}$ as a function of $m_{H^{\pm\pm}}$
for (a) DG1, (b) DG2, (c) DG3 and (d) DG4.
}
\label{fig6}
\end{figure}

\section{Conclusions}

We have studied the production
of doubly charged Higgs bosons ($H^{\pm\pm}$)
at hadron colliders in the Higgs Triplet Model
(HTM), in which a complex $Y=2$
scalar triplet is added to the Standard Model.
The HTM can explain the observed neutrino
mass matrix by invoking 
Yukawa couplings $h_{ij}$ of the triplet fields to leptons.
A definitive signal of the HTM would be the observation
of the decay $H^{\pm\pm}\to l^\pm l^\pm$, which enjoys 
almost negligible background at hadron colliders, and
whose branching ratios are correlated with the neutrino
mass matrix. We studied the production mechanism 
$q'\overline q\to H^{\pm\pm}H^{\mp}$ which can be as
large as the mechanism $q\overline q\to H^{++}H^{--}$
assumed in the current searches at the Tevatron.
Since the present search strategy is sensitive
to single production of $H^{\pm\pm}$,  
we advocated the use of the inclusive single production 
cross-section ($\sigma_{H^{\pm\pm}}$) when comparing
the experimentally excluded region with the 
theoretical cross-section. This leads to a strengthening of 
the mass bound for $m_{H^{\pm\pm}}$ which
now carries a dependence on $m_{H^{\pm}}$, and significantly
improves the $H^{\pm\pm}$ search potential at the Tevatron
and LHC. Although we 
performed our numerical analysis in the HTM,
we emphasized that the introduction 
of $\sigma_{H^{\pm\pm}}$ is also relevant for 
any model which contains a $Y=2$ Higgs triplet 
(e.g. L-R symmetric models and Little Higgs Models).

Moreover, we quantified the impact of the 
decay mode $H^{\pm\pm}\to H^\pm W^*$ for the case
of a hierarchical, inverted hierarchical and 
degenerate neutrino mass spectrum. On discovering a $H^{\pm\pm}$
it would be imperative to measure the  
absolute value of $h_{ij}$ (and hence $v_{\Delta}$)
in order to reconstruct the low energy Higgs triplet Lagrangian.
We stressed that an order of magnitude estimate
of $h_{ij}$ could be obtained if the channel 
$H^{\pm\pm}\to H^\pm W^*$ is observed {\sl and}
$m_{H^\pm}$ is roughly measured. We encourage
a detailed experimental simulation of this decay mode at
both the Tevatron and LHC.

\end{document}